\documentclass[]{pasj01_nohead}
\usepackage{bm}		
\usepackage{url}
\usepackage{color}

\newcommand{\dpar}[2]{\frac{\partial#1}{\partial#2}}

\newcommand{\bracketfunc}[2]{\left(\frac{#1}{#2}\right)}
\newcommand{\rhosurf}{\Sigma}

\newcommand{\vrad}{\rm{v}_{R}}
\newcommand{\vphi}{\rm{v}_{\phi}}

\newcommand{\mstar}{M_{\ast}}
\newcommand{\mpl}{M_{\rm p}}
\newcommand{\rp}{R_{\rm p}}
\newcommand{\hp}{h_{\rm p}}
\newcommand{\vphiavg}{\vphi^{\rm avg}}
\newcommand{\dvphi}{\delta \vphi}
\newcommand{\fj}{F_{J}}
\newcommand{\fms}{F_{M}}
\newcommand{\tp}{T_{\rm p}}
\newcommand{\tponeside}{T_{\rm p}^{\infty}}
\newcommand{\sigmamin}{\Sigma_{\min}}
\newcommand{\sigmaminp}{\Sigma'_{\min}}
\newcommand{\gapwidth}{\Delta_{\rm gap}}
\newcommand{\sigmaedge}{\Sigma_{\rm th}}
\newcommand{\fjwave}{F_{J}^{\rm wave}}
\newcommand{\fjwavem}{F_{J,m}^{\rm wave}}
\newcommand{\fjvis}{F_{J}^{\rm vis}}
\newcommand{\tdeposit}{T_{\rm d}}

\newcommand{\lambdad}{\Lambda_{\rm d}}
\newcommand{\xdep}{x_{\rm d}}
\newcommand{\wdep}{w_{\rm d}}
\newcommand{\Omegak}{\Omega_{k}}
\newcommand{\Omegakp}{\Omega_{k,p}}
\newcommand{\citepeg}[1]{(e.g., \cite{#1})}

\newcommand{\icarus}{Icarus}

\newcommand{\RED}[1]{#1}

\begin{document} 
\Received{\today}
\Accepted{}

\title{Modelling of deep gaps created by giant planets in protoplanetary disks}

\author{Kazuhiro D. \textsc{Kanagawa}\altaffilmark{1}%
\thanks{}}
\altaffiltext{1}{Institute of Physics and CASA$^{\ast}$, Faculty of Mathematics and Physics, University of Szezecin, Wielkopolska 15, PL-70-451 Szczecin, Poland}
\email{kazuhiro.kanagawa@usz.edu.pl}

\author{Hidekazu \textsc{Tanaka}\altaffilmark{2}}
\altaffiltext{2}{Astronomical Institute, Tohoku University, Sendai, Miyagi 980-8578, Japan}

\author{Takayuki \textsc{Muto}\altaffilmark{3}}
\altaffiltext{3}{Division of Liberal Arts, Kogakuin University, 1-24-2 Nishi-Shinjuku, Shinjuku-ku, Tokyo 163-8677, Japan}

\author{Takayuki \textsc{Tanigawa}\altaffilmark{4}}
\altaffiltext{3}{National Institute of Technology, Ichinoseki College, Ichinoseki-shi, Iwate 021-8511, Japan}



\maketitle

\begin{abstract}
A giant planet embedded in a protoplanetary disk creates a gap.
This process is important for both theory and observations.
Using results of a survey for a wide parameter range with two-dimensional hydrodynamic simulations, we constructed an empirical formula for the gap structure (i.e., the radial surface density distribution), which can reproduce the gap width and depth obtained by two-dimensional simulations.
This formula enables us to judge whether an observed gap is likely to be caused by an embedded planet or not.
The propagation of waves launched by the planet is closely connected to the gap structure.
It makes the gap wider and shallower as compared with the case where an instantaneous wave damping is assumed.
The hydrodynamic simulations shows that the waves do not decay immediately at the launching point of waves, even when the planet is as massive as Jupiter.
Based on the results of hydrodynamic simulations, we also obtained an empirical model of wave propagation and damping for the cases of deep gaps.
The one-dimensional gap model with our wave propagation model is able to well reproduce the gap structures in hydrodynamic simulations.
In the case of a Jupiter-mass planet, we also found that the waves with smaller wavenumber (e.g., $m=2$) are excited and transport the angular momentum to the location far away from the planet.
The wave with $m=2$ is closely related with a secondary wave launched by the site opposite from the planet.
\end{abstract}

\section{Introduction} \label{sec:intro}
During the past decade, more than a thousand extrasolar planets have been discovered.
A survey of extrasolar planets has revealed the diversity of giant planets outside the solar system \citepeg{Burke2014}.
Giant planets are born in protoplanetary disks, due to core accretion \citepeg{Mizuno1980,Kanagawa_Fujimoto2013} or to the gravitational instability of the gaseous disk \citepeg{Cameron1978,Zhu2012b}.
Once formed, they undergo orbital migration, and gas accretion results in growth their mass.
The diversity of extrasolar planets is closely connected to such processes \citepeg{Mordasini_Alibert_Benz_Klahr_henning2012,Ida_Lin_Nagasawa2013}.

Planet forming regions in protoplanetary disks can now be directly imaged, such as by the Atacama Long Millimeter/Submillimeter Array (ALMA) and by eight-metre class optical and/or near-infrared telescopes.
High-resolution images have revealed the presence of complex morphological structures in disks; these structures include spirals \citepeg{Muto2012,Grady2013, Christiaens_Casassus_Perez_VanderPlas_Menard2014,Benisty2015,Currie2015,Akiyama2016} and gaps \citepeg{Osorio2014,ALMA_HLTau2015,Momose2015,Akiyama2015,Nomura_etal2016, HLTau_HCO2016,Tsukagoshi2016}.
Direct images are now finding possible signatures of planets being formed in disks \citepeg{Sallum2015}.
To understand the origins of the disk structures and their possible connection to the formation of planets, it is important to construct appropriate quantitative models.

A planet interacts gravitationally with the gas in the surrounding disk, and as a result, 
the planet excites density waves (spirals).
If the planet is sufficiently massive, it also creates gap structures \citepeg{Lin_Papaloizou1979,Goldreich_Tremaine1980,Artymowicz_Lubow1994,Kley1999,Crida_Morbidelli_Masset2006}.
Recently there have been a number of studies on the quantitative relationship between the mass of a planet and the gap structure (i.e., depth and width) \citep{Duffell_MacFadyen2013,Fung_Shi_Chiang2014,Kanagawa2015b,Duffell_Chiang2015,Fung_Chiang2016,Kanagawa2016a}, and the application to actual observations has been discussed \citep{Kanagawa2015b,Momose2015,Marel2016,Nomura_etal2016,Tsukagoshi2016}.
However, there is still room for improvement of the current models.
\cite{Kanagawa2016a} presented an empirical formula for the gap width: they defined it to be the location at which the gap surface density is depleted by $50\%$ from its original value.
The overall structure of the gap created by the planet should be further investigated.

The gap structure is closely associated with the damping of spiral density waves excited by a planet, because the planet exchanges the angular momentum to the disk via the waves \citepeg{Takeuchi_Miyama_Lin1996,Goodman_Rafikov2001,Rafikov2002a,Dong2011}.
\cite{Duffell2015} constructed analytical models of the gap structure by using the wave damping model of \cite{Goodman_Rafikov2001}, which is particularly useful in the case of shallow gaps (created by relatively low-mass planets).
However, the wave propagation in the case where the deep gap is formed by a high-mass planet is still poorly understood.
The propagation of waves induced by the high-mass planet would be qualitatively different from that in the case of the low-mass planet, as implied by the parametrized study by \cite{Kanagawa2015a} (hereafter, K15).

In this paper, we extend our previous model of the gap width and depth, and present a complete model of the gap shape; this is based on a number of two-dimensional long-term hydrodynamic simulations.
We also provide a model of the wave propagation indicated by the hydrodynamic simulations, when the deep gap is formed.
In Section~\ref{sec:basic_eq}, we briefly summarize the setup of the numerical simulations we performed.
In Section~\ref{sec:results}, we obtain an empirical formula for the gap structure, showing the results of hydrodynamic simulations.
From this empirical formula of the gap structure, we can build an empirical model of propagation of the density waves.
In Section~\ref{sec:1dmodel}, we obtain the empirical model of the wave propagation.
Adopting this model of the wave propagation, we provide a semianalytical radial one-dimensional model of the gap structure which is able to reproduce the gap structures given by the two-dimensional simulations.
\RED{
In Section~\ref{sec:discussion}, we discuss an observational application of our formula.
We also discuss how the wave excitation and propagation are changed along with an increase of a planet mass, showing the results of the two-dimensional hydrodynamic simulations.
A discussion about hydrodynamical instabilities such as the Rayleigh instability and Rossby wave instability is included in this section.
}
Section~\ref{sec:summary} contains a summary and discussion.

\section{Summary of setup of numerical simulations} \label{sec:basic_eq}
\cite{Kanagawa2016a} performed a number of the two-dimensional simulations over a wide range of parameter space (planet mass, disk scale height and viscosity), and derived a relationship between the gap width and the planet mass.
In this paper, we extend the model by further analyzing of the numerical simulations.
In this section, we briefly review the setup of hydrodynamic simulations.

We numerically calculated the gap formation processes in a protoplanetary disk in the presence of a planet.
We assumed a geometrically thin and non-self-gravitating disk.
We adopted a two-dimensional cylindrical coordinate system ($R,\phi$), and the origin was located at the position of the central star.
We adopted a simple locally isothermal equation of state, and the distribution of the temperature is independent of time.
The mass of the central star is also constant during the simulations.
Using FARGO \citep{Masset2000}, which is widely used in the study of disk--planet interaction \citepeg{Crida_Morbidelli2007,Baruteau_Meru_Paardekooper2011,Zhu2011}, we solved the equations of continuity and motion of disk gas.
In this paper, we adopt $\alpha$-prescription of \cite{Shakura_Sunyaev1973}, and then the kinetic viscosity is written as $\nu=\alpha h^2 \Omegak^{-1}$, where $h$ and $\Omegak$ are the disk scale height and the Keplerian angular velocity, respectively.

The initial disk structures, boundary conditions and other computational setup (e.g., computational domain and resolutions) are as the same as those in \cite{Kanagawa2016a}.
To complete the parameter survey, we additionally ran 6 simulations with $\hp/\rp = 2/15$, where $\hp$ is the disk scale height at $\rp$ which is the orbital radius of the planet; ($\mpl/\mstar$,$\alpha$) = ($10^{-3},4\times 10^{-3}$), ($2\times 10^{-3},4\times 10^{-3}$), ($5\times 10^{-4},10^{-3}$),($10^{-3},10^{-3}$), ($5\times 10^{-4},6.4\times 10^{-4}$), and ($10^{-3},6.4\times 10^{-4}$), and we also ran 2 simulations with small planet mass of $\mpl/\mstar=5\times 10^{-5}$, where $\mpl$ and $\mstar$ are the masses of the planet and the central star, respectively; $(\alpha$, $\hp/\rp$) = ($10^{-3}$, $1/20$) and ($10^{-3}$,$1/25$) (totally 34 runs).
\RED{The survey covers the range of the planet mass as $5\times 10^{-5} < \mpl/\mstar < 2\times 10^{-3}$, the range of the disk aspect ratio as $1/30 < \hp/\rp < 2/15$, and the range of the disk viscosity as $10^{-4}<\alpha < 10^{-2}$.
The parameter set of each run is described in \cite{Kanagawa2016a}, except above additional runs.
}

The gap is opened by the disk--planet interaction and is closed by the viscous diffusion.
In steady state, the viscous angular momentum flux is balanced by the planetary torque.
The timescale of the gap opening would be scaled by the viscous timescale \citepeg{Lynden-Bell_Pringle1974}.
As an empirical formula, the gap width ($\gapwidth$), which is defined as a width of a region where the surface density is smaller than $0.5 \Sigma_0$ in steady state, where $\Sigma_0$ is the surface density outside the gap, is obtained as follows \citep{Kanagawa2016a}:
\begin{eqnarray}
	\frac{\gapwidth}{\rp} &= 0.41 K'^{1/4},
	\label{eq:gapwidth_0.5}
\end{eqnarray}
where
\begin{eqnarray}
	K'&=\left(\frac{\mpl}{\mstar} \right)^2 \left( \frac{\hp}{\rp} \right)^{-3} \alpha^{-1}.
	\label{eq:kp}
\end{eqnarray}
The timescale of the gap opening ($t_{\rm vis}$) is
\begin{eqnarray}
	t_{\rm vis}= \left(\frac{\gapwidth}{2\rp} \right)^2 \left( \frac{\hp}{\rp} \right)^{-2} \alpha^{-1} \Omega_p^{-1}.
	\label{eq:vistime_gap}
\end{eqnarray}
Using equation~(\ref{eq:gapwidth_0.5}), this timescale is obtained as
\begin{eqnarray}
	t_{\rm vis} &= 0.24 \bracketfunc{\mpl/\mstar}{10^{-3}} \bracketfunc{\hp/\rp}{0.05}^{-7/2} \bracketfunc{\alpha}{10^{-3}}^{-3/2} \nonumber \\
	& \qquad \qquad \times \bracketfunc{\mstar}{1M_{\odot}}^{-1/2}\bracketfunc{\rp}{10\rm{AU}}^{3/2} \mbox{Myr}.
	\label{eq:gap_vistime}
\end{eqnarray}
To obtain the steady state, we have to calculate the evolution until $t\sim t_{\rm vis}$ (see, Appendix~\ref{sec:timeevo_various}).
Note that for nominal parameters, this timescale is usually shorter than or comparable with the migration timescale of the type II ($\sim R_p^2/\nu $) or disk lifetime ($\sim 1$Myr).
Hence, a full-width gap can be observed.

\section{Empirical formula for gap structures} \label{sec:results}
\begin{figure*}
	\begin{center}
		\resizebox{0.98\textwidth}{!}{\includegraphics{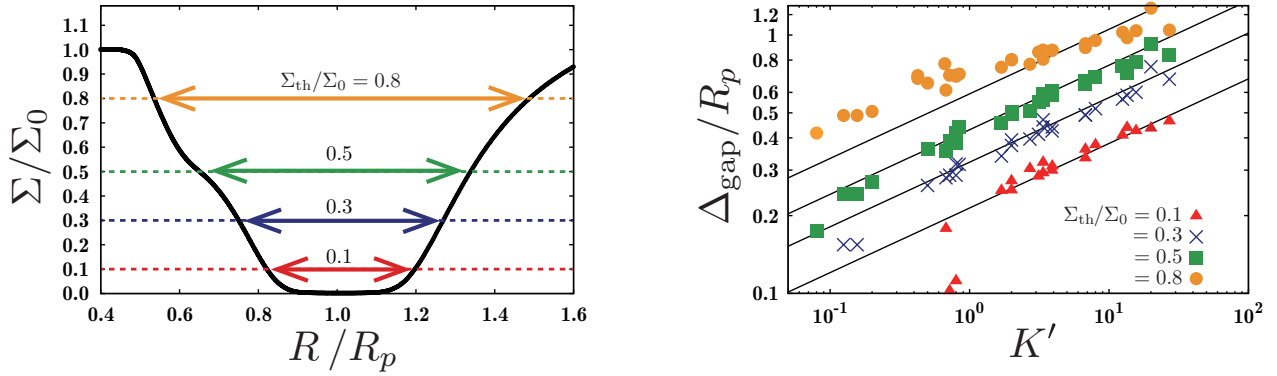}}
		\caption{
		{(\textit{Left})} Schematic picture of measurement of the gap widths with various $\sigmaedge$.
		{(\textit{Right})} The gap widths with $\sigmaedge =0.1\Sigma_0, 0.3\Sigma_0, 0.5\Sigma_0,$ and $0.8\Sigma_0$.
		The thin lines show the results for equation~(\ref{eq:gapwidth_varsedge}) with $\sigmaedge=0.1\Sigma_0, 0.3\Sigma_0, 0.5\Sigma_0,$ and $0.8\Sigma_0$ (from the bottom).
		\label{fig:gapwidths}
		}
	\end{center}
\end{figure*}
\RED{
In this section, we obtain an empirical formula for radial structure of the planet-induced gap in steady state.
\cite{Kanagawa2016a} have found the scaling relation of the gap width, which is defined by a radial width of a region where the surface density is smaller than a density threshold of $\sigmaedge = 0.5\Sigma_0$.
First we here show that this scaling relation can be extended even when the density threshold is different from $0.5\Sigma_0$.
As illustrated in the left panel of Figure~\ref{fig:gapwidths}, we measure the gap widths for $\sigmaedge=0.1\Sigma_0$, $0.3\Sigma_0$, $0.5\Sigma_0$, and $0.8\Sigma_0$.
The right panel of Figure~\ref{fig:gapwidths} shows the gap width with each $\sigmaedge$ in terms of $K'$.
As can be seen in the figure, the gap widths defined by $\sigmaedge=0.1\Sigma_0$ and $0.3\Sigma_0$ are proportional to $K'^{1/4}$, as was seen with $\sigmaedge=0.5\Sigma_0$.
Note that we confirmed that the widths shown in the figure reach to the steady values at $t\simeq t_{\rm vis}$ given by equation~(\ref{eq:vistime_gap}) (see, Appendix~\ref{sec:timeevo_various}).
Although the gap widths defined by $\sigmaedge=0.8\Sigma_0$ deviate slightly from the line defined by $\propto K'^{1/4}$ for $K'$  $\lesssim 1$, they are roughly proportional to $K'^{1/4}$.
We obtain an scaling relation including that shown by \cite{Kanagawa2016a} as,
\begin{eqnarray}
	\frac{\gapwidth \left( \sigmaedge \right)}{\rp} &= \left( 0.5 \frac{\sigmaedge}{\Sigma_0} + 0.16 \right)K'^{1/4}.
	\label{eq:gapwidth_varsedge}
\end{eqnarray}
This scaling relation agrees with the results given by previous hydrodynamic simulations (see, Appendix~\ref{sec:width_comp_previous_hydro}).
Note that the above relationship corresponds to equation~(\ref{eq:gapwidth_0.5}) when $\sigmaedge /\Sigma_0 = 0.5$.
Equation~(\ref{eq:gapwidth_varsedge}) is in good agreement with the gap widths shown in Figure~\ref{fig:gapwidths}.

\begin{figure*}
	\begin{center}
		\resizebox{0.98\textwidth}{!}{\includegraphics{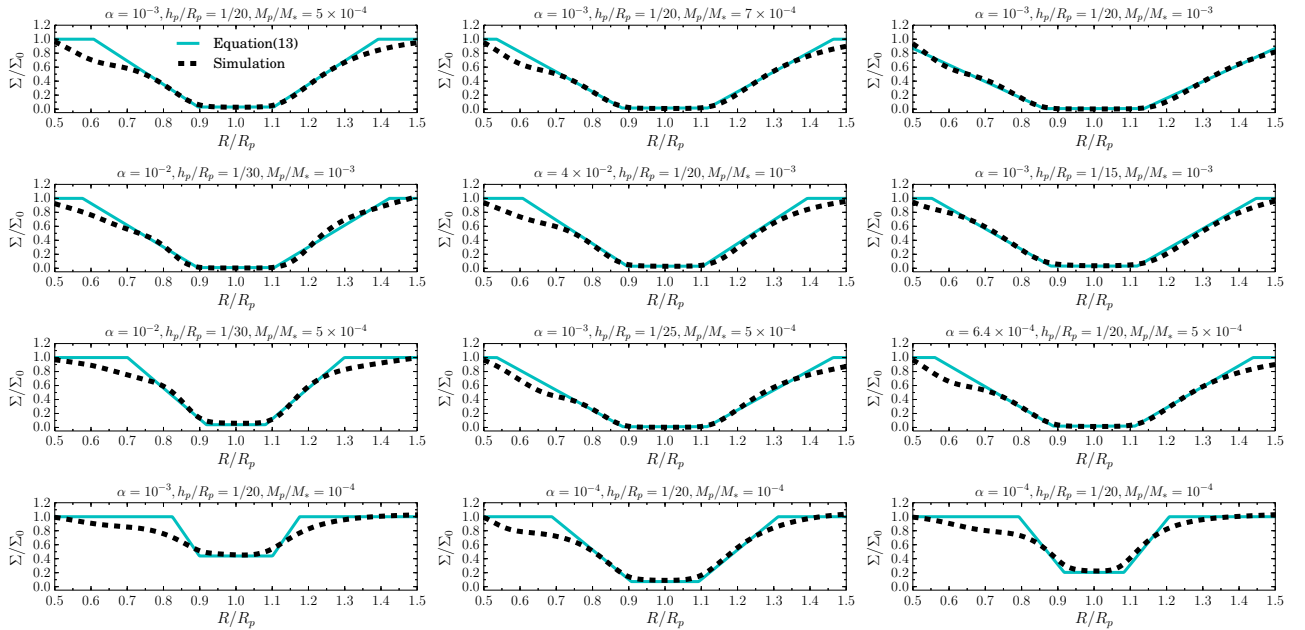}}
		\caption{
		Azimuthal averaged surface densities given by the empirical formula of equation~(\ref{eq:gap}) (solid line) and the hydrodynamic simulations (dashed line) for various sets of planet mass, disk aspect ratio, and viscosity.
		\label{fig:avgdens_simple}
		}
	\end{center}
\end{figure*}
We can derive a formula of a radial distribution of the gas density from equation~(\ref{eq:gapwidth_varsedge}), setting $\sigmaedge=\Sigma(R)$ and $\gapwidth = 2|R-\rp|$ as,
\begin{eqnarray}
\Sigma(R) &= \left\{
\begin{array}{ll}
\Sigma_{\min}       &\ \mbox{for}\ \ |R-R_p|< \Delta R_1,\\
\Sigma_{\rm gap}(R) &\ \mbox{for}\ \ \Delta R_1< |R-R_p| < \Delta R_2,\\
\Sigma_0            &\ \mbox{for}\ \ |R-R_p|> \Delta R_2,
\end{array}\right.
\label{eq:gap}
\end{eqnarray}
with
\begin{eqnarray}
	\frac{\Sigma_{\rm gap} (R)}{\Sigma_0} &= 4.0K'^{-1/4} \frac{|R-\rp|}{\rp} - 0.32,
	\label{eq:sigma_gapedge}
\end{eqnarray}
where $\Delta R_1$ and $\Delta R_2$ are given by
\begin{eqnarray}
	\Delta R_1 &= \left(\frac{\sigmamin}{4\Sigma_0} + 0.08 \right) K'^{1/4} R_p \label{eq:r1}, \\
	\Delta R_2 &= 0.33 K'^{1/4} R_p \label{eq:r2}.
\end{eqnarray}
The surface density of the gap bottom $\Sigma_{\min}$ is obtained by  \citepeg{Duffell_MacFadyen2013,Fung_Shi_Chiang2014,Kanagawa2015a}
\begin{eqnarray}
	\frac{\sigmamin}{\Sigma_0} &= \frac{1}{1+0.04K}, 
	\label{eq:smin}
\end{eqnarray}
where
\begin{eqnarray}
	K&=\left(\frac{\mpl}{\mstar} \right)^2 \left( \frac{\hp}{\rp} \right)^{-5} \alpha^{-1}.
	\label{eq:k}
\end{eqnarray}
We illustrate the surface density given by the empirical formula of equation~(\ref{eq:gap}) along with results of hydrodynamic simulations in Figure~\ref{fig:avgdens_simple}.
In the top row of the figure, we show the gap structures of $\mpl/\mstar=5\times 10^{-4}$ to $10^{-3}$ for $\hp/\rp=1/20$ and $\alpha=10^{-3}$.
For this case, equation~(\ref{eq:gap}) accurately reproduces the gap structures obtained by the two-dimensional hydrodynamic simulations.
The second and third rows show the gap structures of $\mpl/\mstar=10^{-3}$ and $5\times 10^{-4}$ for various disk aspect ratios and viscosities.
As in the top row, equation~(\ref{eq:gap}) accurately reproduces the results of the hydrodynamic simulations.
The bottom row shows the gap structures of $\mpl/\mstar=10^{-4}$.
For $(\alpha, \hp/\rp) = (10^{-4},1/20)$ and $(10^{-3},1/25)$, equation~(\ref{eq:gap}) is still able to fit the gap structure of the simulations, but for $(\alpha,\hp/\rp) = (10^{-3},1/20)$, the gap obtained by the simulation is wider than that given by the equation.
Equation~(\ref{eq:gap}) does not give a good estimate when the gap is shallow, that is, $\sigmamin \gtrsim 0.5 \Sigma_0$.
}

\begin{figure*}
	\begin{center}
		\resizebox{0.98\textwidth}{!}{\includegraphics{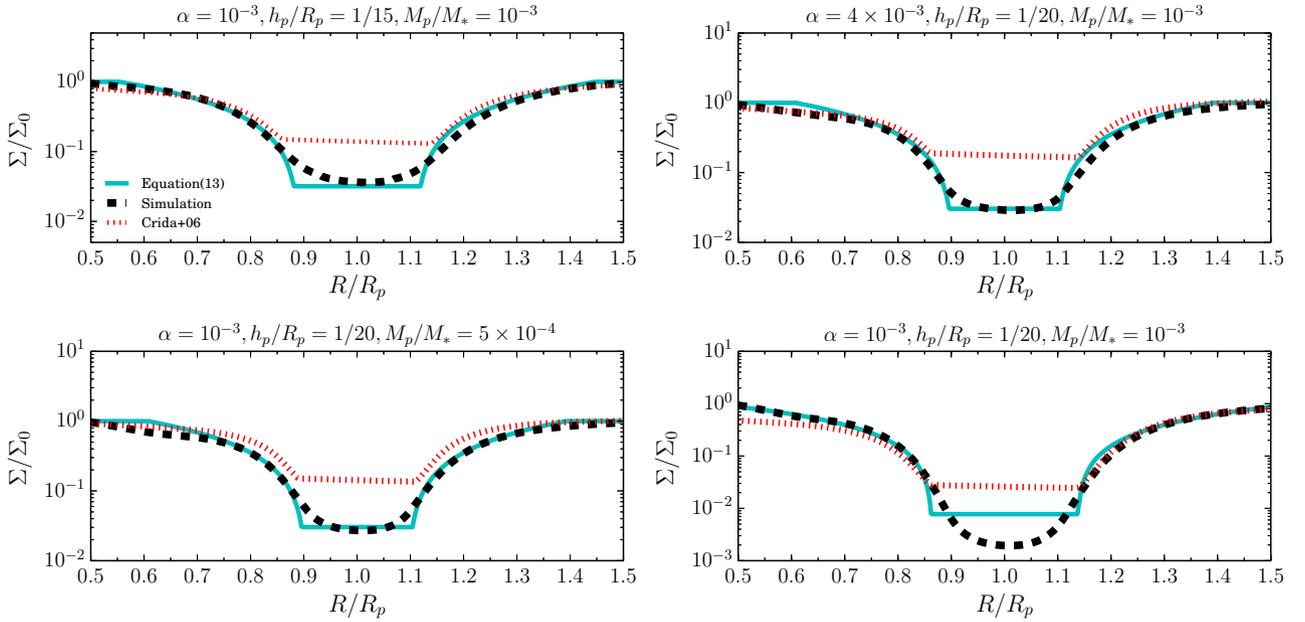}}
		\caption{
		Comparison between equation~(\ref{eq:gap}) (solid line) and the model of \protect \cite{Crida_Morbidelli_Masset2006} (dotted line).
		The dashed lines are the results of two-dimensional hydrodynamic simulations.
		\label{fig:comp_crida_fitmodel}
		}
	\end{center}
\end{figure*}
\cite{Crida_Morbidelli_Masset2006} provided a semianalytical model for deep gaps.
They introduced ``pressure torque'' and considered the balance between the pressure torque, the planetary torque and the viscous angular momentum flux.
In Figure~\ref{fig:comp_crida_fitmodel}, we compare gap radial structures given by equation~(\ref{eq:gap}) and \cite{Crida_Morbidelli_Masset2006}.
We obtain Crida's model in the figure by solving equation~(14) of \cite{Crida_Morbidelli_Masset2006}.
In the calculation of Crida's model, we adopt 1.5 times larger $\Sigma_0$ than ours, to set $\rhosurf$ outside the gap in Crida's model to be the same level as that of our model.
Crida's models result in significantly shallower gaps compared to the gaps obtained in the two-dimensional simulations with our parameter set.
Equation~(\ref{eq:gap}) nicely reproduces the gap depths in the case of the parameters presented in the bottom right panel of Figure~\ref{fig:comp_crida_fitmodel}.
In the case with $\alpha=10^{-3}$, $\hp/\rp=1/20$ and $\mpl/\mstar=10^{-3}$ (the bottom right case), the gap depth obtained by equation~(\ref{eq:gap}) is about 5 times shallower than that in the simulation.
This discrepancy partially comes from fact that the gas is not rotating at Kepler velocity.
We provide a semianalytical model that takes into account this effect in Section~\ref{sec:1dmodel}.
For the gap width, both equation~(\ref{eq:gap}) and Crida's model reasonably agree with the two-dimensional simulations.
However, when the planet mass is small, Crida's model gives narrower gaps than the simulation, as can be seen from the bottom  left panel in Figure~\ref{fig:comp_crida_fitmodel}.
\section{Semianalytical model of radial gap structures} \label{sec:1dmodel}
\subsection{Model of wave propagation} \label{sec:model_wavepropagation}
\RED{
A planet exchanges angular momentum with the surrounding disk gas through density waves.
As pointed out by previous studies \citepeg{Petrovich_Rafikov2012,Kanagawa2015a}, a propagation of the density waves is directly connected with the surface density distribution of the gap.
The planet gives angular momentum to density waves.
Then, the angular momentum from the planet is carried by the waves.
Finally the angular momentum is deposited to the disk gas by dissipation of the waves \citepeg{Takeuchi_Miyama_Lin1996,Goodman_Rafikov2001}.
In this section, considering the wave propagation which is suggested by our empirical formula of the surface density distributions of the gap obtained in the previous section, we present a one-dimensional model of the gap.
}

From the conservation of the angular momentum,
\begin{eqnarray}
	R \vphiavg \fms  + \fjvis (R)= \fj(\rp) + \tdeposit(R),
	\label{eq:flux2}	
\end{eqnarray}
where $\vphiavg$ is the azimuthal averaged $\vphi$ and $\fj(\rp)$ is the angular momentum flux at the radius of the orbit of the planet \citep{Kanagawa2015a}.
The viscous angular momentum flux $\fjvis$ is given by equation~(\ref{eq:amfvis}) in Appendix~\ref{sec:model_description}.
Furthermore, the cumulative torque deposited to the disk, $\tdeposit(R)$ is defined by equation~(\ref{eq:def_deposited_amf}), which is equal to the difference between the cumulative torque excited by the planet, $\tp(R)$, and the angular momentum flux due to the density waves, $\fjwave(R)$ which is defined by equation~\ref{eq:amfwave}. 
\begin{figure}
	\begin{center}
		\resizebox{0.49\textwidth}{!}{\includegraphics{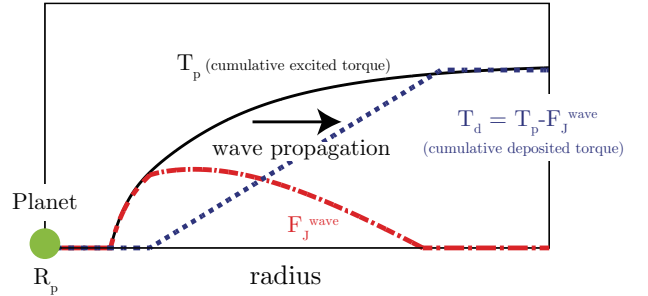}}
	\caption{
		Schematic picture of the angular momentum transfer due to the density waves.
		The wave propagation makes difference between the cumulative planetary torque $\tp$ and the cumulative deposited torque $\tdeposit$.
		The difference between them corresponds to the angular momentum flux due to the waves $\fjwave$.
		See Appendix~\ref{sec:model_description} for detail.
		\label{fig:schematic_wavepropagation}
		}
	\end{center}
\end{figure}
In Figure~\ref{fig:schematic_wavepropagation}, we show a schematic picture of the effect of the wave propagation.
(see \cite{Kanagawa2015a} and Appendix~\ref{sec:model_description} for detail).

\RED{
The surface density is given by equation~(\ref{eq:flux2}).
If assuming that the gap width is smaller than $\rp$, we neglect the terms which is proportional to a power of $\hp/\rp$, as e.g., $R = \rp$ and $\Omega = \Omegakp$, where the suffix $p$ indicates values at $\rp$ in the following.
The viscous angular momentum flux is given by $\fjvis \simeq 2\pi \rp^2 \nu \rhosurf (d\Omega/dR)$.
The derivative of the angular velocity is $d\Omega/dR \simeq -3 \Omegakp/(2\rp) \left[1- (\hp^2/3)(d^2\ln \rhosurf / dR^2) \right]$, of which second term is not negligible since $d^2\ln \rhosurf /dR^2 \sim 1/\hp^2$ in the gap region \citep{Kanagawa2015a}.
Then, we rewrite equation~(\ref{eq:flux2}) as,
\begin{eqnarray}
&3\pi \rp^2 \nu_p \Omegakp \rhosurf \left[1-\frac{\hp^2}{3} \frac{\rm{d}^2 \ln \rhosurf}{dR^2} \right] \nonumber\\
&\qquad \qquad \qquad \qquad = \fj(\rp)-\rp^2\Omegakp \fms +\tdeposit(R).
\label{eq:flux_width_density}
\end{eqnarray}
As shown in equation~(\ref{eq:flux_width_density}), the surface density distribution is determined if the function of $\tdeposit$ is given.
The function of $\tdeposit$ implicitly depends on the surface density, because $\tdeposit$ depends on the torque exerted by the planet which is determined by the surface density distribution.
To obtain the distributions of the surface density and deposited torque, an iterative method is required (see, \cite{Kanagawa2015a}).
}

The propagation of the waves launched by the planet has been investigated by e.g., \cite{Goodman_Rafikov2001}.
\RED{
Their model indicates that the waves launched by a sufficiently massive planet ($\mpl > M_1 = (2/3)(\hp/\rp)^3\mstar  \simeq 1\times 10^{-4}\mstar$) decay due to the weak shock that develops close to the launching point of the density wave.
Because of the shock, the angular momentum of the waves decreases as $|R-\rp|^{5/4}$ from the shock location.
Using the wave propagation model of \cite{Goodman_Rafikov2001}, \cite{Duffell2015} proposed an analytical model of the gap structure.
His model is able to accurately reproduce the structures of the gap, if the gap is shallow.
When the gap is deep, however, the model of \cite{Duffell2015} predicts a much narrower gap than that obtained by numerical simulations, which implies that the angular momentum deposition from the waves is less effective than that predicted by the model of \cite{Goodman_Rafikov2001}.
}
For the deep gap, \cite{Kanagawa2015a} presented the simple model of wave propagation using two free parameters $\xdep$ and $\wdep$, which represent the position at which waves are damped and the width of the deposition site, respectively, as
\begin{eqnarray}
\tdeposit(R) &= \left\{
\begin{array}{l}
	0 \qquad \quad \mbox{for}\  |R-\rp| < {\displaystyle \xdep-\frac{\wdep}{2}},\\
 	{\displaystyle \frac{\tponeside}{\wdep}} {\displaystyle \left| R-\left(\xdep-\frac{\wdep}{2} \right) \right|} \\
 	\qquad \quad \ \mbox{for} \  {\displaystyle \xdep - \frac{\wdep}{2} < |R-\rp|< \xdep+\frac{\wdep}{2}},\\
 	\tponeside \qquad \mbox{for}\  {\displaystyle |R-\rp| > \xdep+\frac{\wdep}{2}},
\end{array} \right.
\label{eq:deposited_torque}
\end{eqnarray}
where $\tponeside$ is the total excitation torque exerted by the planet (see Appendix~\ref{sec:model_description} for detail).
If we assume that the form of the torque deposit is given by equation~(\ref{eq:deposited_torque}), we can obtain the gap shape using equation~(\ref{eq:flux2}).
It should be noted, however, that equation~(\ref{eq:deposited_torque}) is a purely parametrized model for $\tdeposit(R)$, and its validity should be checked with numerical simulations.
In particular, the gap surface density distribution depends strongly on the choice of the two free parameter $\xdep$ and $\wdep$ and therefore, realistic values for these parameters should be determined using numerical simulations.

Motivated by the surface density distribution of equation~(\ref{eq:gap}), we empirically determine the parameters $\xdep$ and $\wdep$ as
\begin{eqnarray}
	\xdep& = \rp \pm \frac{\Delta R_2+\Delta R_1}{2} \label{eq:xdep},\\
	\wdep& = \Delta R_2-\Delta R_1 \label{eq:wdep},
\end{eqnarray}
where $\Delta R_1$ and $\Delta R_2$ are defined by equations~(\ref{eq:r1}) and (\ref{eq:r2}), respectively, and the sign of $\xdep$ is positive when $R>\rp$ and negative when $R<\rp$.
\begin{figure*}
	\begin{center}
		\resizebox{0.98\textwidth}{!}{\includegraphics{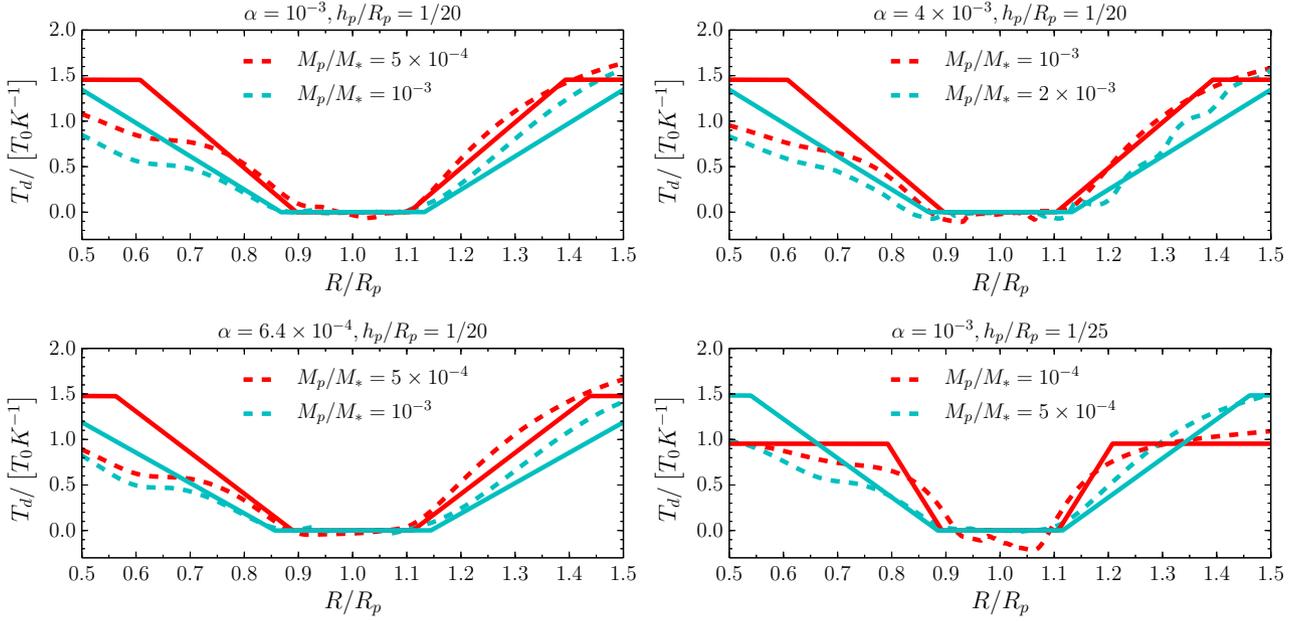}}
	\caption{
		Cumulative torque deposited onto the disk, as obtained by equation~(\ref{eq:deposited_torque})(solid lines) and the hydrodynamic simulations (dashed lines).
		\label{fig:deposited_am_comp_sims_models}
		}
	\end{center}
\end{figure*}
Figure~\ref{fig:deposited_am_comp_sims_models} shows the cumulative torques deposited onto the disk given by equation~(\ref{eq:flux2}) and the two-dimensional hydrodynamic simulations.
The cumulative deposited torque given by equation~(\ref{eq:deposited_torque}) with equations~(\ref{eq:xdep}) and (\ref{eq:wdep}) reasonably reproduces the results of the two-dimensional hydrodynamic simulations.

We use the WKB formula \citepeg{Ward1986} to calculate the excitation torque.
However, we should consider nonlinear effects since we deal with relatively massive planets creating deep gaps in this paper.
According to \cite{Miyoshi_Takeuchi_Tanaka_Ida1999}, the nonlinear effect reduces the excitation torque as compared with that expected by the linear theory.
In order to include the nonlinear effect, we introduce a reduction factor $f_{\rm NL}(<1)$ and multiply this factor by the torque density (see Appendix~\ref{sec:model_description} for detail).
As shown below, $f_{\rm NL} = 0.4$ better reproduces the gap depth obtained by the numerical hydrodynamic simulations.
Since we consider cases where $M_p>(\hp/\rp)^3$, the factor $f_{\rm NL}$ is always set to $0.4$.

\subsection{Radial gap structures} \label{subsec:rhosurf}
\begin{figure*}
	\begin{center}
		\resizebox{0.98\textwidth}{!}{\includegraphics{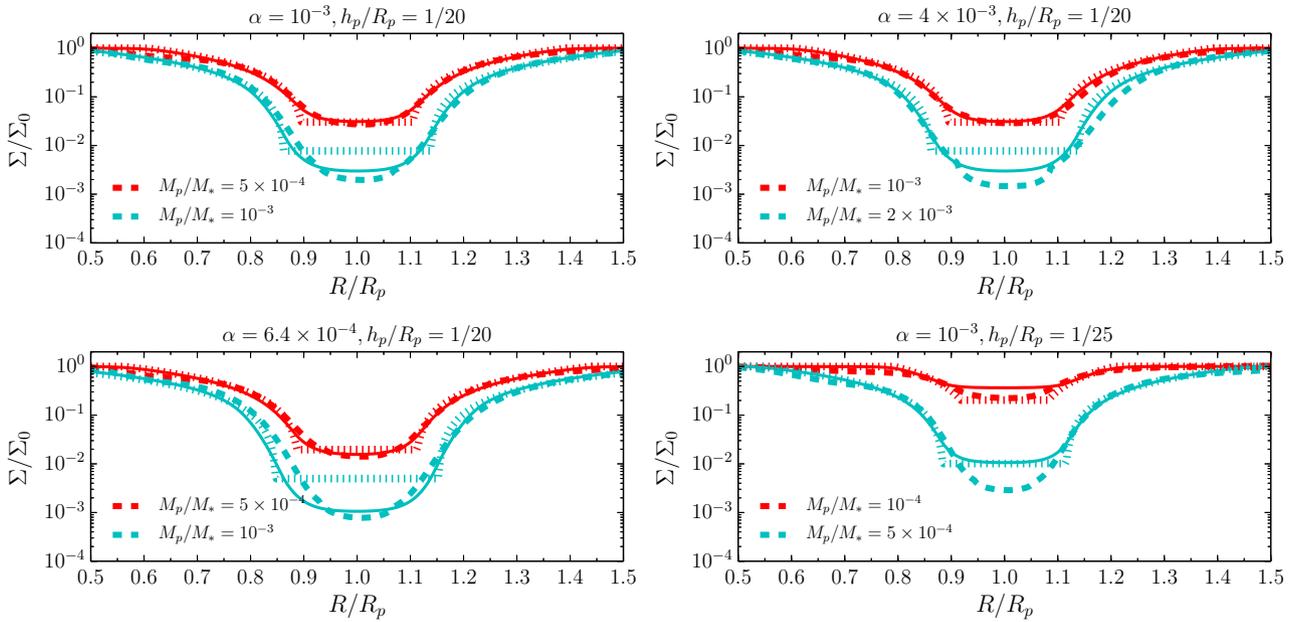}}
	\caption{
 		Surface densities obtained by equation~(\ref{eq:flux2}) (solid lines) and the hydrodynamic simulations (dashed lines).
 		For comparison, we also plot the surface density given by equation~(\ref{eq:gap}) (dotted line).
 		\label{fig:avgdens_comp_sims_models}
		}
	\end{center}
\end{figure*}
In Figure~\ref{fig:avgdens_comp_sims_models}, we show the radial structure of a gap, calculated by solving equation~(\ref{eq:flux_width_density}) with $\xdep$ and $\wdep$ given by equations~(\ref{eq:xdep}) and (\ref{eq:wdep}).
For comparison, we plot gap structures obtained by the two-dimensional hydrodynamic simulations and equation~(\ref{eq:gap}) in the figure.
The solutions of equation~(\ref{eq:flux2}) are in good agreement on the width and depth with the hydrodynamic simulations for various disk scale heights and viscosities, as long as the planet mass is relatively large.
Since the deviation from Keplerian rotation is taken into account in the semianalytical model, the structure of gap bottom is smoother and the gap is slightly deeper than those given by equation~(\ref{eq:gap}).
When $\mpl/\mstar=10^{-3}$, $\alpha=10^{-3}$, and $\hp/\rp=1/25$ (bottom right panel of Figure~\ref{fig:avgdens_comp_sims_models}), the semianalytical model (equation~\ref{eq:flux2}) gives approximately three time shallower gap than that given by the two-dimensional simulation.
We consider that the semianalytical model provides a reasonable fit to deep gaps created by a planet.

\begin{figure*}
	\begin{center}
		\resizebox{0.98\textwidth}{!}{\includegraphics{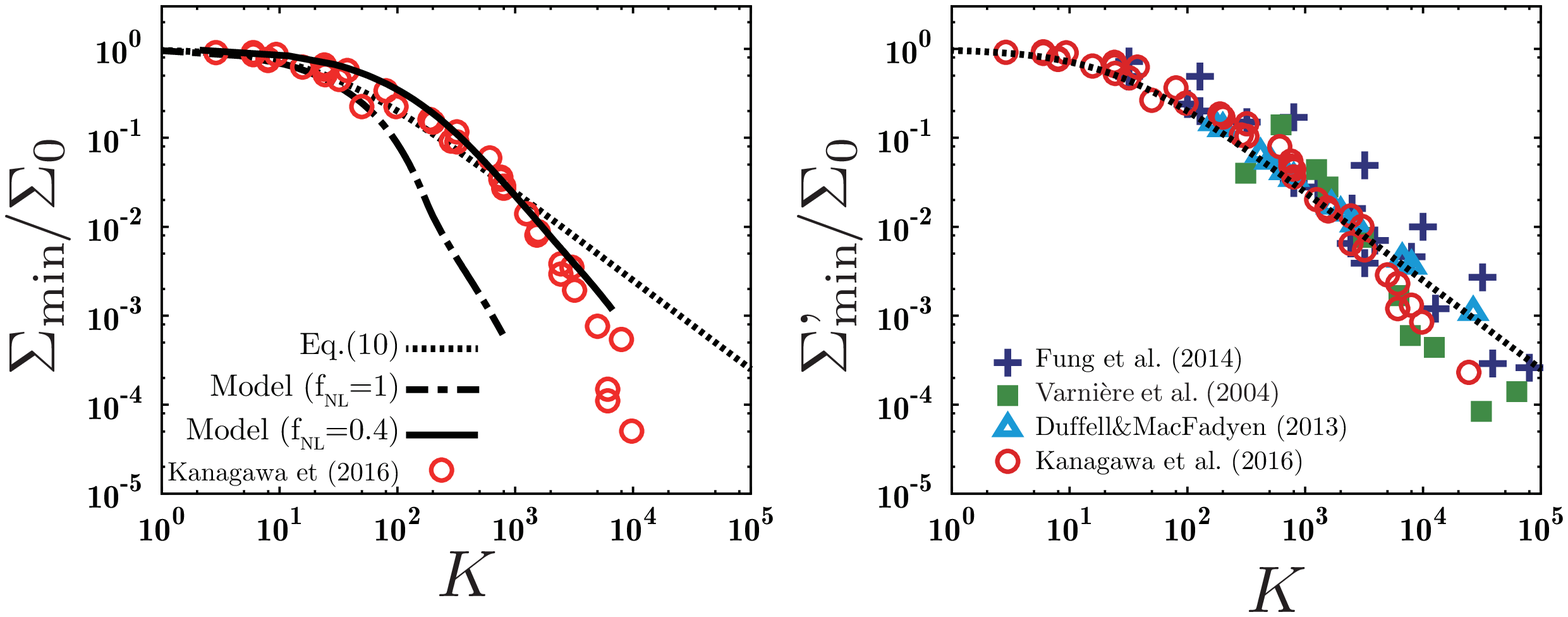}}
	\caption{
		({\textit{Left}}) Minimum value of the azimuthally averaged surface density $\sigmamin$ versus the parameter $K$.
		The circles indicate $\sigmamin$ obtained from the two-dimensional simulaitons by \protect \cite{Kanagawa2016a}.
		The dotted line shows equation~(\ref{eq:smin}), and the solid and chain lines indicate $\sigmamin$, as given by the solution of equation~(\ref{eq:flux2}) with $f_{\rm NL}=0.4$ and $1.0$, respectively.
		({\textit{Right}}) Surface density averaged throughout the gap bottom $\sigmaminp$, as given by \protect \cite{Kanagawa2016a} (circles).
		The crosses, squares, and triangles are the gap depths given by \protect \cite{Fung_Shi_Chiang2014}, \protect \cite{Varniere_Quillen_Frank2004}, and \protect \cite{Duffell_Chiang2015}, respectively.
		\label{fig:gapdepth_comp}
		}
	\end{center}
\end{figure*}
\RED{
The left panel of Figure~\ref{fig:gapdepth_comp} shows the minimum surface density of the gap $\sigmamin$ obtained by the one-dimensional model with $f_{\rm NL}=0.4$ and $1.0$.
We also plot the minimum value of the azimuthal averaged surface density given by the two-dimensional simulations.
When the gap is shallow as $K<10^{2}$, the one-dimensional model (equation~\ref{eq:flux_width_density}) gives us a similar $\sigmamin$ as that given by the two-dimensional simulations, which are also consistent with equation~(\ref{eq:smin}), in both the cases of $f_{\rm NL}=0.4$ and $1$ (but the model with $f_{\rm NL} =1$ looks more appropriate).
As the gap is deeper in $K>10^{2}$, the one-dimensional model with $f_{\rm NL}=1$ gives us a much smaller $\sigmamin$ as compared with equation~(\ref{eq:smin}) (and 2D simulations).
On the other hand, $\sigmamin$ obtained by the one-dimensional model are consistent with these given by two-dimensional simulations and equation~(\ref{eq:smin}).
For very deep gap with $K>10^{3}$, $\sigmamin$ given by two-dimensional simulations is smaller than that predicted by the empirical formula of equation~(\ref{eq:smin}).
Even in this case, the one-dimensional model with $f_{\rm NL} = 0.4$ partially fits $\sigmamin$ given by the simulations.
The assumption of $f_{\rm NL}=0.4$ is reasonably acceptable in the cases with $K<10^{4}$.
When the gap is very deep, $\sigmamin$ very sensitively depends on the value of $\tponeside$.
More sophisticated model of wave excitation by a giant planet would be required to reproduce the depth of the very deep gaps with $K>10^{4}$.
 

The empirical formula of equation~(\ref{eq:smin}) predicts a larger value of $\sigmamin$ than that given by the two-dimensional simulations for $K>10^{3}$, as shown above.
However, equation~(\ref{eq:smin}) is still useful for $\sigmaminp$ which is the surface density averaged throughout the gap bottom region\footnote{The surface density is averaged over the annulus spanning $R=\rp-\delta$ to $\rp+\delta$ with $\delta \equiv 2\rm{max}(R_H,\rp)$, excised from $\theta = \theta_p-\delta/\rp$ to $\theta_p+\delta/\rp$.
This density is similar to the density of the gap bottom which \cite{Fung_Shi_Chiang2014} used (we have not taken time-average of the density).}, as \cite{Fung_Shi_Chiang2014} showed.
In the right panel of Figure~\ref{fig:gapdepth_comp}, we illustrate $\sigmaminp$ given by the two-dimensional simulations.
When the gap is not very deep ($K\lesssim 10^{3}$), both $\sigmaminp$ and $\sigmamin$ are reproduced by equation~(\ref{eq:smin}).
Even when the gap is very deep with $K\gtrsim 10^{3}$, $\sigmaminp$ is also reasonably consistent with equation~(\ref{eq:smin}), whereas $\sigmamin$ is much smaller than that as shown in the left panel of the figure.
The value of $\sigmaminp$ represents the surface density around $R=\rp \pm 2R_H$, rather than $\sigmamin=\rhosurf(\rp)$.
According to \cite{Tanigawa_Watanabe2002}, the gas accretion rate onto the planet depends on the surface density around the location that is twice the Hill radius apart from the planet.
Hence, equation~(\ref{eq:smin}) may be useful to estimate the gas accretion rate, whereas it overestimates the minimum surface density of the gap when $K\gtrsim 10^{3}$.
}

\section{Discussion} \label{sec:discussion}
\subsection{Observational applications} \label{subsec:obs_applications}
\subsubsection{Constraint from the gap radial structure} \label{sss:obs_width_relation}
\begin{figure}
	\begin{center}
		\resizebox{0.49\textwidth}{!}{\includegraphics{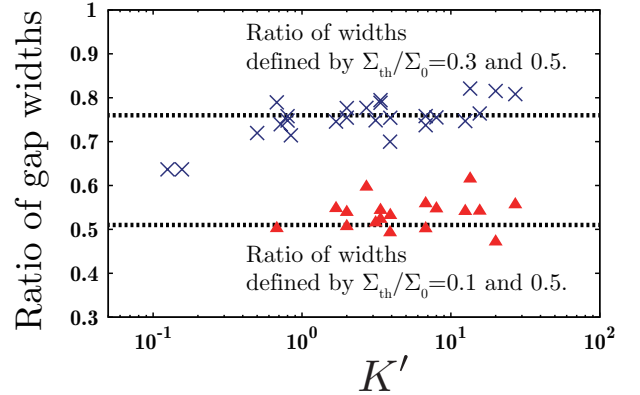}}
	\caption{
	The ratio of the gap width defined by $\sigmaedge=0.1\Sigma_0$ to that defined by $\sigmaedge=0.5\Sigma_0$ (triangles) and $\sigmaedge=0.3\Sigma_0$ and $0.5\Sigma_0$ (crosses).
	The horizontal dotted lines indicate the values predicted by equation~(\ref{eq:ratio_widths}).
		\label{fig:gapwidth_ratio}
		}
	\end{center}
\end{figure}
Here we would like to discuss observational applications of our empirical formula of the gap structure of equation~(\ref{eq:gap}).
Recently, many gap structures have been discovered by ALMA; these have been found not only in dust \citepeg{ALMA_HLTau2015} but also in gas \citepeg{HLTau_HCO2016}.
Such gap structures in protoplanetary disks can be created by dust growth \citep{Zhang_Blake_Bergin2015}, sintering \citep{Okuzumi_Momose_Sirono_Kobayashi_Tanaka2016}, the effects of MRI \citep{Flock_Ruge_Dzyurkevich_Henning_Klahr_Wolf2015}, secular gravitational instability due to gas--dust friction \citep{Takahashi_Inutsuka2016a}, and disk--planet interaction.
It is difficult to distinguish the origin from observations.
However, using the relationship of the gap radial structure that is shown above, we may be able to assess whether the observed gap is due to the disk--planet interaction.

Using equation~(\ref{eq:gapwidth_varsedge}), we can determine the ratio of the gap width for any two arbitrary surface densities $\Sigma_{\rm th,a}$ and $\Sigma_{\rm th,b}$ as 
\begin{eqnarray}
	\frac{\Delta_{\rm gap}(\Sigma_{\rm th,a})}{\Delta_{\rm gap}(\Sigma_{\rm th,b})} = \frac{\Sigma_{\rm th,a}/{\Sigma_0}+0.32}{\Sigma_{\rm th,b}/{\Sigma_0}+0.32}.
	\label{eq:ratio_widths} 
\end{eqnarray}
As can be seen from the above equation, the ratio of the gap widths depends only on the surface densities defined by the widths.
In Figure~\ref{fig:gapwidth_ratio}, we show two ratios of gap widths, defined by $\sigmaedge=0.1\Sigma_0$ and $0.5 \Sigma_0$, and by $\sigmaedge=0.3\Sigma_0$ and $0.5 \Sigma_0$.
Using equation~(\ref{fig:gapwidth_ratio}), we obtain that the ratios are $0.51$ and $0.76$, respectively.
As can be seen in the figure, the ratios obtained from the simulations are reasonably consistent with those obtained with equation~(\ref{eq:ratio_widths}).

\subsubsection{Gap depth--width relation} \label{sss:obs_depth-width_relation}
As shown in \cite{Kanagawa2016a}, there is a relationship between the depth and width of a gap.
Combining equations~(\ref{eq:gapwidth_varsedge}) and (\ref{eq:smin}), we can derive this relationship  in terms of the locations of $\sigmaedge$, as follows:
\begin{eqnarray}
	&\frac{\Delta_{\rm gap} \left(\sigmaedge /\Sigma_0 \right)}{R_p} \left(\frac{\sigmaminp}{\Sigma_0 - \sigmaminp} \right)^{1/4} \left(\frac{\hp}{\rp}\right)^{-1/2} \nonumber\\
	&\qquad \qquad \qquad \qquad \qquad \qquad \qquad = 1.16\frac{\sigmaedge}{\Sigma_0} + 0.35.
	\label{eq:rel_depth_width}
\end{eqnarray}
If the gap is created by a planet, its width, as measured by the above-surface density, should  satisfy equation~(\ref{eq:rel_depth_width}).
When the gap structure is completely resolved and the disk aspect ratio is precisely estimated, we can use the gap widths measured by the different surface density at the gap edge and equation~(\ref{eq:rel_depth_width}) to strictly judge whether the gap was created by the planet.

\subsubsection{Applicability of the model} \label{sss:obs_caveats}
We now discuss some considerations when equations~(\ref{eq:ratio_widths}) and (\ref{eq:rel_depth_width}) are applied to observations, even if the observed gap is sufficiently resolved.
First, we must consider that the distribution of dust particles may be different from that of a gas when the relations are applied to an observation of dust thermal emissions.
Because of dust filtration \citepeg{Zhu2012,Dipierro_Price_Laibe_Hirsh_Cerioli_Lodato2015,Picogna_Kley2015,Rosotti_Juhasz_Booth_Clarke2016}, a gap in dust can be deeper and wider than in gas if the size of the dust particles (or the Stokes number of particles) is relatively large.
Hence, equations~(\ref{eq:ratio_widths}) and (\ref{eq:rel_depth_width}) should be used for observations of disk gas.
Alternatively, the dust--gas coupling depends on the gas surface density, as well as on the size of the dust particles.
If the gas density is sufficiently large, the gas and dust particles will be well mixed.
In this case, equations~(\ref{eq:ratio_widths}) and (\ref{eq:rel_depth_width}) would provide a good estimate.

Second, we assume that the gap structure is in steady state.
As shown in Appendix~\ref{sec:timeevo_various}, the timescale required for the gap to be in steady state can be roughly estimated as $t_{\rm vis} \sim 0.1$ Myr.
If an observed gap is younger than this timescale, the gap will be narrower than it would be in steady state.
Such young gaps cannot be estimated by equations~(\ref{eq:ratio_widths}) and (\ref{eq:rel_depth_width}).

We note that equations~(\ref{eq:ratio_widths}) and (\ref{eq:rel_depth_width}) should be used for relatively deep gaps.
As shown in Figure~\ref{fig:gapwidths}, for a shallow gap ($K'<1$), the actual gap width measured for a larger surface density is slightly wider than that estimated by equation~(\ref{eq:rel_depth_width}).
Because of this, in this case, equation~(\ref{eq:rel_depth_width}) would estimate the disk scale height to be about $1.5$ times the actual value.
Moreover, as also discussed in \cite{Kanagawa2016a}, the observational uncertainties of the orbital radius of the planet and aspect ratio should be taken into account.

Finally, we briefly make comments about our assumption of two-dimensional disks and spatially constant kinematic viscosity $\nu$.
\cite{Fung_Chiang2016} performed three-dimensional simulations of disk--planet interaction in the case when a deep gap is induced by a planet.
The gap profile (i.e., the depth and the width) is not very different from the results of two-dimensional calculations.
Hence, our models may be valid even when three-dimensional effects are properly taken into account.
The assumption of constant viscosity may not be always satisfied.
\cite{Zhu_Stone_Rafikov2013} performed ideal MHD simulations of the interaction between a low-mass planet and a disk and showed that the effective viscosity $\alpha$ within the gap is approximately twice as large as that outside the gap region.
Therefore, the gap may be shallower than our model.
However, since the dependence of $\alpha$ on the gap depth is not very strong (see, equation~\ref{eq:smin}), this effect may not significantly influence our results.

\subsection{Excitation and propagation of waves with deep gaps} \label{subsec:waveprop}
\subsubsection{Planetary torque and angular momentum deposition in the disk} \label{subsec:wave_2d}
In Section~\ref{sec:1dmodel}, we present the empirical model of wave propagation and obtain the semianalytical one-dimentional model of gap structure using this wave propagation model.
\RED{
In the previous studies \citepeg{Goodman_Rafikov2001}, when the planet mass is sufficiently large as $\mpl> M_1 (\sim 1 \times 10^{-4}\mstar \mbox{ if }\hp/\rp = 1/20$), the angular momentum flux of the wave decreases quickly ($\fjwave \propto |R-\rp|^{-5/4}$) from the location near the launching point.
However, our model of the wave propagation (equations~\ref{eq:deposited_torque},\ref{eq:xdep}, and \ref{eq:wdep}) implies that the wave propagation is different from that expected by the model of \cite{Goodman_Rafikov2001}, when the planet is large.
}
For a better understanding of wave propagation with a deep gap, we discuss how the wave propagation properties change as we increase the planet mass.

\begin{figure*}
	\begin{center}
		\resizebox{0.98\textwidth}{!}{\includegraphics{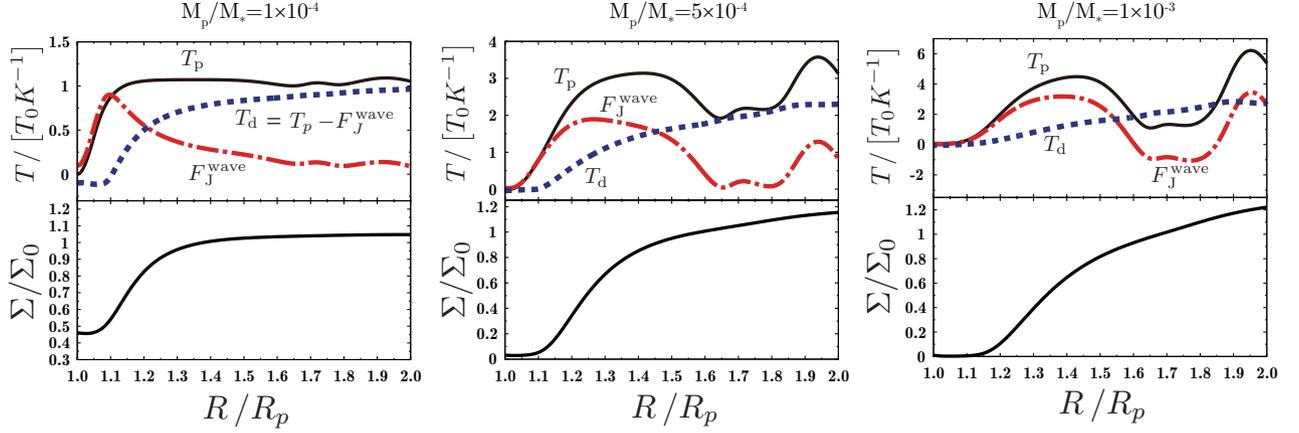}}
	\caption{
		({\textit Top}) Cumulative planetary torque ($\tp$, solid line), angular momentum flux due to the waves ($\fjwave$, chain line), and cumulative deposited torque ($\tdeposit$, dashed line) for $\mpl/\mstar=10^{-4}$ (left), $5\times 10^{-4}$ (middle), and $10^{-3}$ (right).
		The disk aspect ratio and viscosity were set to $1/20$ and $10^{-3}$, respectively.
		({\textit Bottom}) Azimuthal averaged surface density distribution in each case.
		\label{fig:torque_waves_in_simulation}
		}
	\end{center}
\end{figure*}
\RED{
In Figure~\ref{fig:torque_waves_in_simulation}, we show the cumulative planetary torque $\tp$, the angular momentum flux due to the waves $\fjwave$, and the cumulative deposited torque $\tdeposit$ given by two-dimensional hydrodynamic simulations, for a small planet ($\mpl/\mstar=10^{-4}$), a planet with moderate mass ($\mpl/\mstar = 5\times 10^{-4}$), and a giant planet ($\mpl/\mstar = 10^{-3}$).
The disk aspect ratio and the viscosity are $1/20$ and $10^{-3}$, respectively.
In the case of $\mpl/\mstar=10^{-4}$ (left panel of Figure~\ref{fig:torque_waves_in_simulation}), the wave excitation is in the linear regime. 
In this case, $\tp$ increases within the gap bottom region of $\rhosurf \simeq \sigmamin$, and it is saturated to be constant value outside the gap.
The angular momentum flux of the waves $\fjwave$ also increases with the cumulative planetary torque when $R/\rp \lesssim 1.1$.
For $R/\rp>1.1$, $\fjwave$ decreases while $\tp$ does not change, which indicates that the waves are damping due to the shock and deposit the angular momentum to the disk.
The angular momentum flux of the waves $\fjwave$ is approximately proportional to $|R-\rp|^{-5/4}$ and quickly decreases within the density gap ($R/\rp \lesssim 1.3$).
With decreasing $\fjwave$, the cumulative deposited torque $\tdeposit$ increases.
Because $\fjwave=0$ outside of the gap, $\tdeposit$ converges to $\tp$ at the location apart from the planet.

When $\mpl/\mstar=5\times10^{-4}$ (the middle panel in Figure~\ref{fig:torque_waves_in_simulation}), the evolution of the waves should be in the nonlinear regime.
In this case, the cumulative planetary torque increases not only at the gap bottom but also at the gap edge of $\rhosurf \sim \Sigma_0$.
The angular momentum flux of the waves is almost equal to $\tp$ when $R/\rp < 1.15$.
For $R/\rp>1.15$, $\fjwave$ deviates from $\tp$ and decreases due to the shock.
As compared with the case of $\mpl/\mstar=10^{-4}$, the decrease of $\fjwave$ is much slow, which indicates the angular momentum deposition of the waves is less effective as the planet mass increases.
Outside of the gap, the behavior of $\tp$ is significantly different from that in the case of $\mpl/\mstar=10^{-4}$.
Around $R/\rp = 1.5$, $\tp$ decreases and then it increases again around $R/\rp=1.8$, which indicates that the planet interacts with the gas at such the large distance from the planet.
The angular momentum flux of the waves $\fjwave$ also changes along with $\tp$.
The cumulative deposited torque $\tdeposit$ slowly increases in this region.
In the right panel of Figure~\ref{fig:torque_waves_in_simulation}, the planet mass is $\mpl/\mstar=10^{-3}$, and the excitation of waves is highly nonlinear.
The behaviors of $\tp$, $\fjwave$, and $\tdeposit$ are similar to these in the case of $\mpl/\mstar=5\times 10^{-4}$, but the decrease of $\fjwave$ is further slow, and then the increase of $\tdeposit$ is also very slow.
}

\subsubsection{Excitation and damping of waves in low-m modes} \label{subsec:wave_2d}
To further examine the wave propagation when a deep gap is present, we investigate the wave resonances associated with small wavenumber using the results of two-dimensional simulations.
We consider the Fourier components for $\rhosurf$, $\vrad$, $\vphi$, and $\Psi$ which are given by
\begin{eqnarray}
	f_m(R,m)&=\frac{1}{2\pi} \int^{2\pi}_{0} f(R,\phi) \exp\left( -im\phi \right) d\phi,
	\label{eq:fourier_trans}
\end{eqnarray}
where $f$ is $\rhosurf$, $\vrad$, $\vphi$, or $\Psi$.
For convenience, we will let the subscript $m$ indicate the $m$-th Fourier component.
The torque density exerted on the $m$-th resonance is given by \citep{Goldreich_Tremaine1980}
\begin{eqnarray}
	\left( \frac{d\tp}{dR} \right)_m &=4\pi R \Re(\Psi_m) \Im(\rhosurf_m),
	\label{eq:tqdens_m}
\end{eqnarray}
and the integrated torque exerted on the $m$-th resonance is given by
\begin{eqnarray}
	\tp{}_{,m}&=\int^{R}_{\rp} \left( \frac{d\tp}{dR} \right)_{m} dR=\int^{R}_{\rp}4\pi R \Re(\Psi_m) \Im(\rhosurf_m) dR.
	\label{eq:int_torque_m}
\end{eqnarray}
The angular momentum flux doe to the $m$-th mode of the wave is 
\begin{eqnarray}
	&\fjwavem = \qquad \qquad \qquad \qquad \qquad \qquad \quad \qquad \qquad \qquad \nonumber\\
	&\qquad 4\pi R^2 \left[ \Re( \rhosurf \vrad{}_{,m}) \Re(\delta \vphi{}_{,m}) + \Im( \rhosurf \vrad{}_{,m}) \Im(\delta \vphi{}_{,m}) \right].
	\label{eq:amfm}
\end{eqnarray}
The cumulative torque deposited by the $m$-th mode $\tdeposit{}_{,m}$ can be written as
\begin{eqnarray}
	\tdeposit{}_{,m} &= \tp{}_{,m} - \fjwavem.
\end{eqnarray}

\begin{figure*}
	\begin{center}
		\resizebox{0.98\textwidth}{!}{\includegraphics{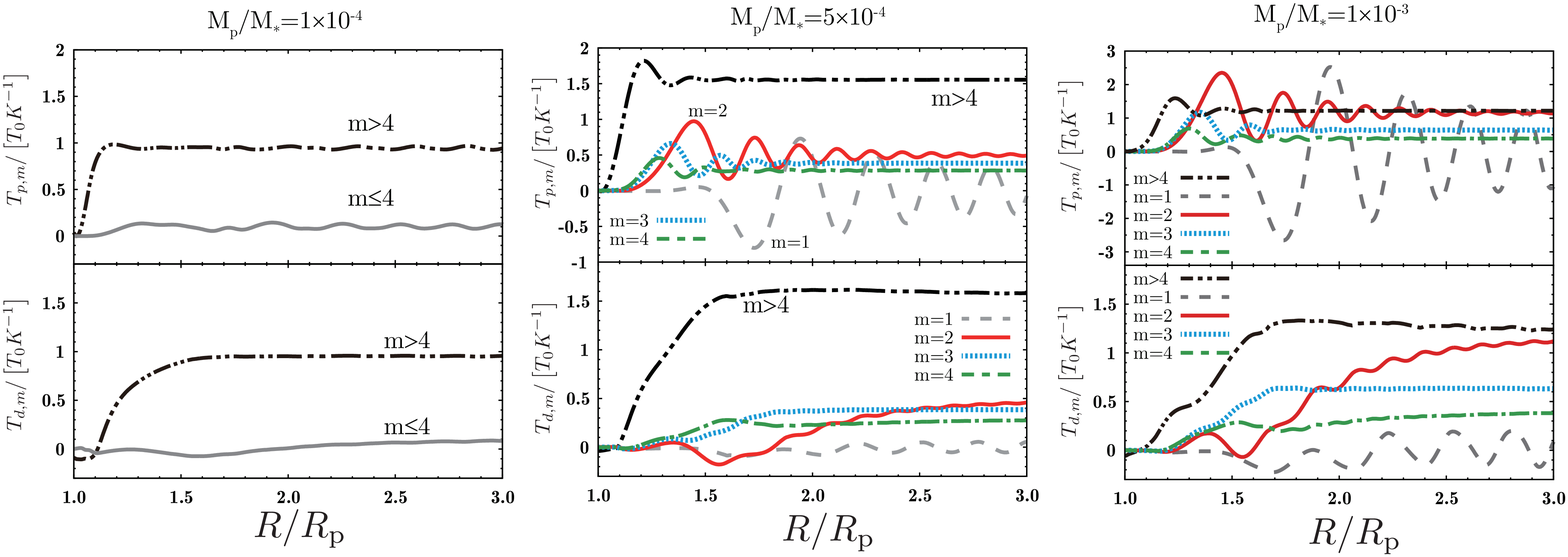}}
	\caption{
		Contributions of the cumulative planetary torque (top) and the cumulative deposited torque (bottom) from azimuthal modes in the same cases as Figure~\ref{fig:torque_waves_in_simulation}.
		The dashed, solid, dotted and dot-dashed double-dot-dashed lines denotes the contributions from $m=1$, $2$,$3$, $4$, and a sum of $m>4$, respectively.
		Only when $\mpl/\mstar=10^{-4}$, we only plot the sum of $m\leq 4$ (gray solid line) and $m>4$, because the contributions from the lower azimuthal modes are very small.
		\label{fig:smallmode}
		}
	\end{center}
\end{figure*}
In Figure~\ref{fig:smallmode}, we illustrate the cumulative torque exerted by the planet and that deposited on the disk for low-m modes.
The parameters for the disk and the planet are the same as those in Figure~\ref{fig:torque_waves_in_simulation}. 
We plot the contributions from $m=1$ -- $4$ individually and the sum of contributions from modes larger than $4$.
In the case of $\mpl/\mstar=10^{-4}$ (left panel of the figure), the planetary torque is mainly exerted by the resonances with $m>4$ since the wave excitation is in the linear regime \citep{Goldreich_Tremaine1980}.
The cumulative deposited torques are also dominated by the contribution from $m>4$ waves.

\RED{
In the cases where the plant mass is large as in the middle and right panels of Figure~\ref{fig:smallmode}, the contributions from the small $m$ resonances are significant.
For instance, in the case of $\mpl/\mstar=10^{-3}$ (the right panel of the figure), the contributions of torque from the resonances of $1\leq m \leq 4$ are comparable to the sum of the contributions from $m>4$ resonances.
The contributions from $m=3,4$ resonances are excited around the gap edge ($R/\rp \simeq 1.2$).
The contributions from $m=1,2$ resonances are significantly excited in the outside of the gap.
The variation of $\tp$ at large distances pointed out in previous subsection (and shown in Figure~\ref{fig:avgdens_comp_sims_models}) is originated from the contributions from $m=1,2$ resonances.
Note that the contribution of $m=1$ resonance is related with the indirect term of the gravitational potential (which is connected with the gravitational interaction between the planet and the central star), which just oscillates around zero at larger distances from the planet, and the contribution of $m=1$ does not significantly influence the angular momentum flux of the waves.

As in the planetary torque, the contributions of the cumulative deposited torque from the small $m$ modes become larger as the planet mass increases.
When $m=2$, in particular, even at radii very far away from the planet ($R/\rp \sim 2$), the cumulative deposited torque slowly increases, in the case of $\mpl/\mstar=10^{-3}$.
These behaviors of $\tp$ and $\tdeposit$ indicate that the waves with small $m$ modes (except $m=1$) significantly excites and carry the angular momentum to the large radii from the planet, with the increase of the planet mass.
As the gap becomes deeper and wider, the contributions on the wave excitation from large $m$ resonances (near the planet) becomes smaller because the surface density at the resonance decreases.
In consequence, the wave excitation on smaller $m$ resonances becomes significant, as discussed by \cite{Juhasz_Benisty_Pohl_Dullemond_Dominik_Paardekooper2015}.
Moreover, \cite{Lee2016} has shown that the excitation of waves with small azimuthal wavenumber can be enhanced due to nonlinear effects.
The significantly excitation of waves with small $m$ would be explained by the gap opening and the nonlinear effects.

We should note that the wave with $m=2$ deposit the angular momentum outside the gap with $\rhosurf > \Sigma_0$ ($R/\rp > 1.8$).
This indicates that wider gap is formed by the deposition of the angular momentum of $m=2$ wave, whereas opening such the wide gap requires very long time that is comparable with the disk life time (see equation~\ref{eq:vistime_gap}).
}

\begin{figure*}
	\begin{center}
		\resizebox{0.98\textwidth}{!}{\includegraphics{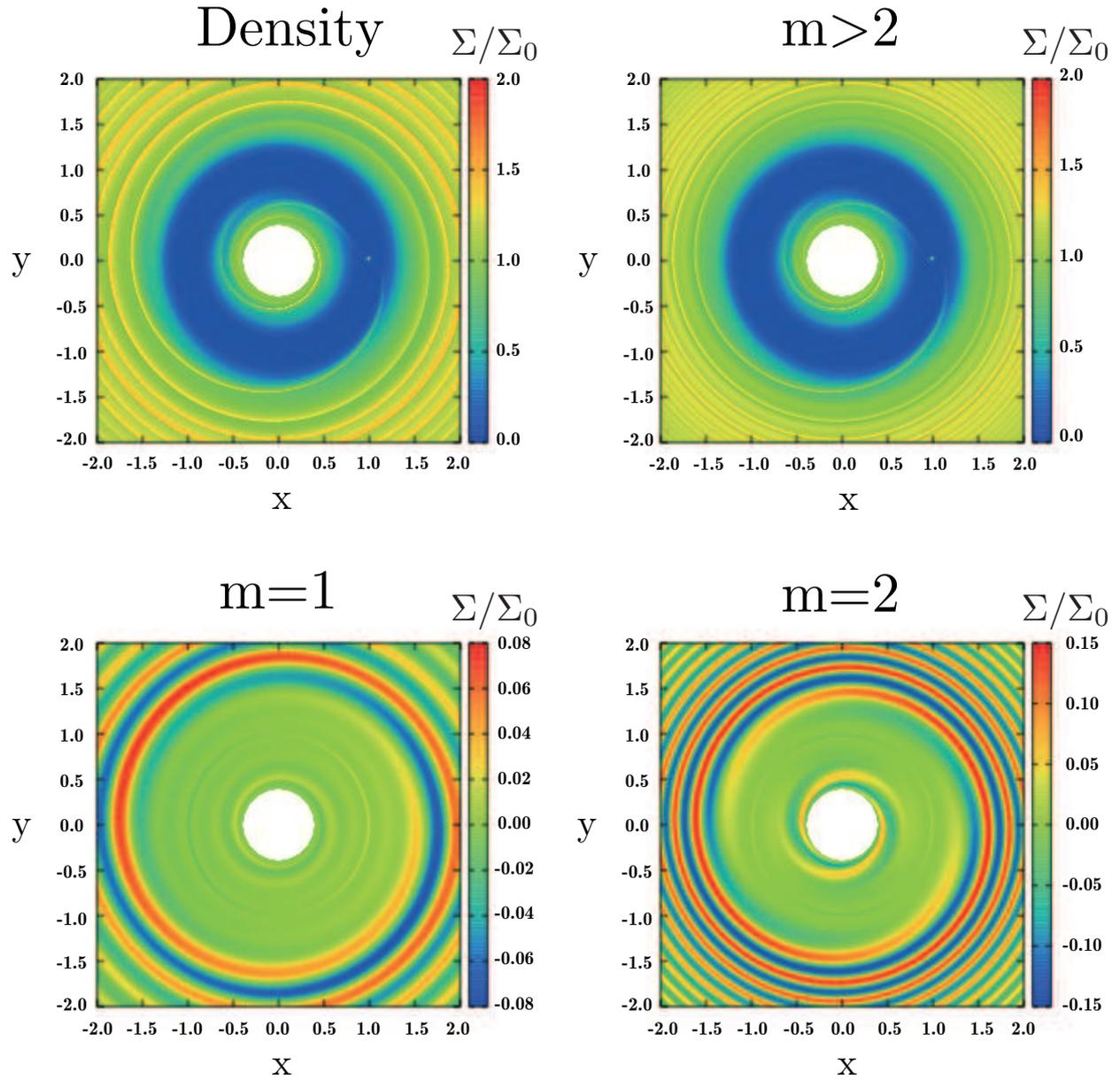}}
	\caption{
	Two-dimensional surface density distribution in the case of $M_p/M_{\ast}=10^{-3}$, $h/R_p=1/20$, and $\alpha=10^{-3}$.
	Shown are the surface densities produced by all modes (left top), the $m>2$ modes (right top), the $m=1$ mode (left bottom), and the $m=2$ mode (right bottom).
	\label{fig:decomp_densmap}
	}
	\end{center}
\end{figure*}
The development of the $m=1, 2$ contributions significantly changes the morphology of the waves in addition to the transport of angular momentum.
In Figure~\ref{fig:decomp_densmap}, we show the surface density produced by the $m=1,2$ components and the sum of the $m>2$ components, for the same case as shown in Figure~\ref{fig:smallmode}.
The surface density produced by the $m$-th component is
\begin{eqnarray}
	\Sigma (m,R,\phi) &= 2\pi \Sigma_m(R,\phi) \exp\left(im\phi \right),
	\label{eq:sigma_m} 
\end{eqnarray}
where $\Sigma_m$ is the Fourier component of the surface density.
The waves that are far from the planet are created by the contributions from the $m=1$ and $m=2$ resonances.
In particular, the secondary wave that is launched from the site opposite the planet is mainly composed of the $m=1$ and $m=2$ components.
Because the contribution from the $m=2$ component is strong, the secondary wave originates at a site opposite from the planet.
\cite{Fung_Dong2015} have shown that the azimuthal separation of the primary and secondary waves depends on the planet mass.
For a large planet, as shown in Figure~\ref{fig:decomp_densmap}, this separation is close to $180^{\circ}$, which is consistent with our results.
For a smaller planet, the separation is smaller than $180^{\circ}$.
In this case, the contributions from other lower modes (e.g., $m=3,4$), are not much smaller than these from $m=1,2$ (see Figure~\ref{fig:smallmode}).
Because of these contributions, the launching point of the secondary wave would move toward the planet.

It is worth noting that the secondary wave interacts with the planet.
As seen in Figure~\ref{fig:decomp_densmap}, the secondary wave passes through the location at which $\phi=\phi_p$ around $R/\rp=1.4$.
Near this location, the planetary torque due to the $m=2$ resonance increases and then decreases, as shown in the top panel of Figure~\ref{fig:smallmode}; this is a result of the interaction between the planet and the secondary wave.
The cumulative deposited torque of $m=2$ also increases and decreases when the secondary wave passes through the location at which $\phi=\phi_p$; this is also a consequence of the interaction between the planet and the secondary wave, and especially of the $m=2$ component.

\RED{
\subsection{Effect of instability for gap structures} \label{subsec:rbw}
Previous studies \citepeg{Li2000,Tanigawa_Ikoma2007,Lin2012,Lin2014,Ono_Nomura_Takeuchi2014,Kanagawa2015a,Ono2016} pointed out a possibility that hydrodynamic instabilities (i.e., Rayleigh Instability, Rossby Wave Instability) occur at the edge of the gap induced by the planet.
The condition for the onset of the RI is given by $d(R^2\Omega)/dR < 0$ \citep{Chandrasekhar1961}.
In our parameter range, however, this condition is not satisfied because the gap structure is less steep.
For a very deep gap with a massive planet as $\mpl/\mstar \simeq 4\times 10^{-3}$, the RI may be violated as shown by \cite{Fung_Chiang2016}.

\begin{figure*}
	\begin{center}
		\resizebox{0.98\textwidth}{!}{\includegraphics{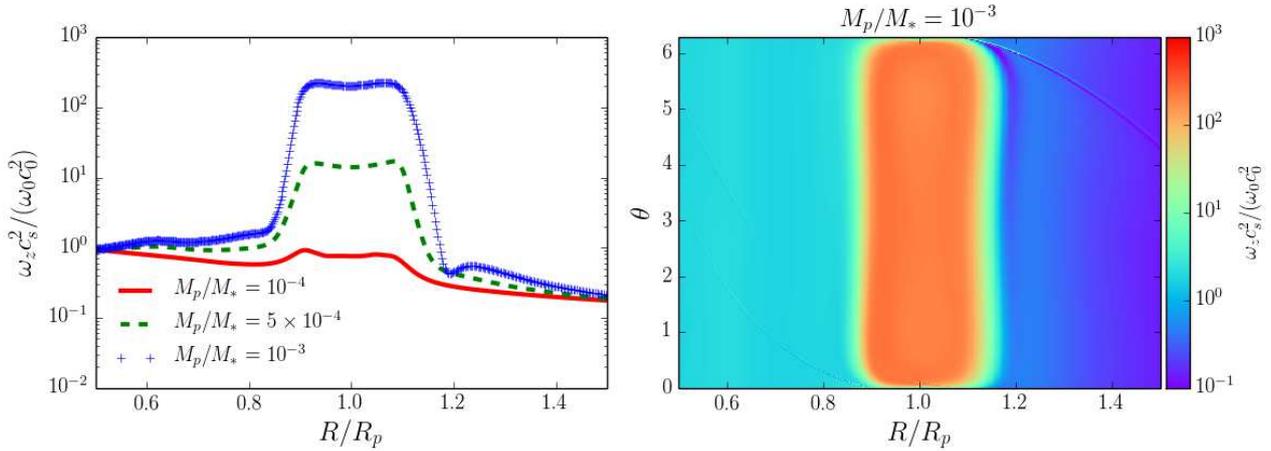}}
	\caption{
	(Left)Distributions of the azimuthal averaged values of $\omega_z c_s^2$ for $\mpl/\mstar = 10^{-4}$ (solid), $5\times 10^{-4}$ (dotted), and $10^{-3}$ (cross), respectively, when $\hp/\rp=1/20$ and $\alpha=10^{-3}$.
	(Right)Two-dimensional distribution of $\omega_z c_s^2$ in the case of $\mpl/\mstar=10^{-3}$ in the left panel.
	\label{fig:vorticities_h0.05}
	}
	\end{center}
\end{figure*}
According to \cite{Lovelace_Colgate_Nelson1999}, the RWI can occur when the function of $\omega_z S^{2/\Gamma} = \omega_z c_s^2$ (when locally isothermal), where $\omega_z = (\nabla \times \bm{V})_z/\rhosurf$ is the potential vorticity and $c_s$ is sound speed, respectively, has an extreme value (a maximum or minimum), whereas \cite{Ono2016} have shown that Lovelace's condition is not a sufficient one.
The left panel of Figure~\ref{fig:vorticities_h0.05} shows radial distributions of azimuthal averaged values of $\omega_z c_s^2$ for $\mpl/\mstar=10^{-4}$, $5\times 10^{-4}$, and $10^{-3}$ when $\hp/\rp=1/20$ and $\alpha=10^{-3}$.
When $\mpl/\mstar=10^{-4}$ and $5\times 10^{-4}$, there is no extreme around the gap edge.
When $\mpl/\mstar=10^{-3}$, we can find a local minimum and maximum around $R/\rp=1.2$.
Even in this case, the two-dimensional distribution of $\omega_z c_s^2$ is almost axis-symmetric (see the right panel of the figure) and there is no vortex-like structure, it does not seem that the RWI affects the gap structure.

The RWI can occur at the edge of the planet-induce gap in the case of the low viscosity (e.g.,\cite{Yu2010,Fu_Li_Lubow_Li2014,Zhu_Stone_Rafikov_Bai2014},see also Appendix~\ref{sec:RWI_inviscid}).
The RWI may constraint the gap structure in this case, as stated by \cite{Hallam_Paardekooper2017}.
However, with viscosity larger than $\alpha \sim 10^{-3}$ -- $10^{-4}$, the RWI is stabilized by the viscosity during the long-term evolution \citepeg{Lin2014,Fu_Li_Lubow_Li2014,Zhu_Stone2014}.
In fact, there is no clear evidence which the RWI affects the gap structure in Figure~\ref{fig:vorticities_h0.05}.
In this case, the effect of the RWI on the gap structure may not be essential.
}

\section{Summary} \label{sec:summary}
Using the results of the survey of the hydrodynamic simulations (34 runs) over the wide parameter space, we derived a quantitative relationship between a planet and the gap radial structure.
We also obtain an empirical model of the wave propagation, from the results of the survey.
Using this empirical model of wave propagation to a semianalytical model of the gap structure provided by \cite{Kanagawa2015a}, we can obtain the gap structures which are in very good agreement with these obtained by the two-dimensional simulations.
Our results can be summarized as follows:
\begin{enumerate}
  \item \RED{We extended the scaling relation of the gap width given by \cite{Kanagawa2016a}.
  From this scaling relation, we derived an empirical formula for the surface density distributions of the gap (equation~\ref{eq:gap}), which accurately reproduces the results of hydrodynamic simulations; see Figure~\ref{fig:avgdens_simple}.}
  \item Our model puts a constraint on the origin of the observed gap structure, as discussed in Section~\ref{subsec:obs_applications}. Using equations~(\ref{eq:ratio_widths}) and (\ref{eq:rel_depth_width}), we can judge whether an observed gap is induced by the planet; see Figure~\ref{fig:gapwidth_ratio}.
  \item We confirmed the validity of the empirical model for wave propagation adopted in \cite{Kanagawa2015a} (equation~\ref{eq:deposited_torque}) with the parameters given by equations~(\ref{eq:xdep}) and (\ref{eq:wdep}).
  This model describes wave propagation that is consistent with the results of the hydrodynamic simulations; see Figure~\ref{fig:deposited_am_comp_sims_models}.
  \RED{Using this wave propagation model, we can reproduce the gap structure obtained by the two-dimensional hydrodynamic simulations by the one-dimensional model; see Figure~\ref{fig:avgdens_comp_sims_models}.}
  \item \RED{Our model of wave propagation indicates that the waves excited by the larger planet carry the angular momentum at larger distances from the planet.}
  As the planet mass is larger, waves with smaller azimuthal wavenumber (e.g.,$m=2$) are excited strongly.
  The angular momentum is transported to the distant locations from the planet, even when the planet mass is larger as Jupiter; see Figure~\ref{fig:smallmode}.
  \item The development of modes with small azimuthal wavenumber changes the morphology of spiral waves.
  When the planet mass is sufficiently large, the $m=2$ mode creates a secondary wave launched from the site opposite from the planet (see Figure~\ref{fig:decomp_densmap}), in addition to a primary wave originated from the location of the planet.
  The secondary wave would be closely related with the transport of angular momentum.
\end{enumerate}

\RED{
As the planet mass increases, the nonlinear effects become significant in excitation and propagation of waves, and thus the mechanism of the gap formation is different from that when the planet is small in linear theory.
}
However, these theoretical mechanisms are not yet completely understood.
To understand the mechanism of the gap formation and the morphology of the density waves induced by a giant planet, it will be necessary to further investigate the wave excitation and propagation when there are deep gaps.

\begin{ack}
This work was supported by JSPS KAKENHI Grant Numbers 23103004, 26103701, 26800106, and 26800229, and the Polish National Science Centre MAESTRO grant DEC- 2012/06/A/ST9/00276.
KDK was supported by the ALMA Japan Research Grant of the NAOJ Chile Observatory, NAOJ-ALMA-0135.
Numerical computations were carried out on the Cray XC30 at the Center for Computational Astrophysics, National Astronomical Observatory of Japan and the Pan-Okhotsk Information System at the Institute of Low Temperature Science, Hokkaido University.
\end{ack}

\appendix 
\section{Time variations of the gap depth and width} \label{sec:timeevo_various}
Here we consider time-variation of the depth and width of gaps.
\begin{figure*}
	\begin{center}
		\resizebox{0.98\textwidth}{!}{\includegraphics{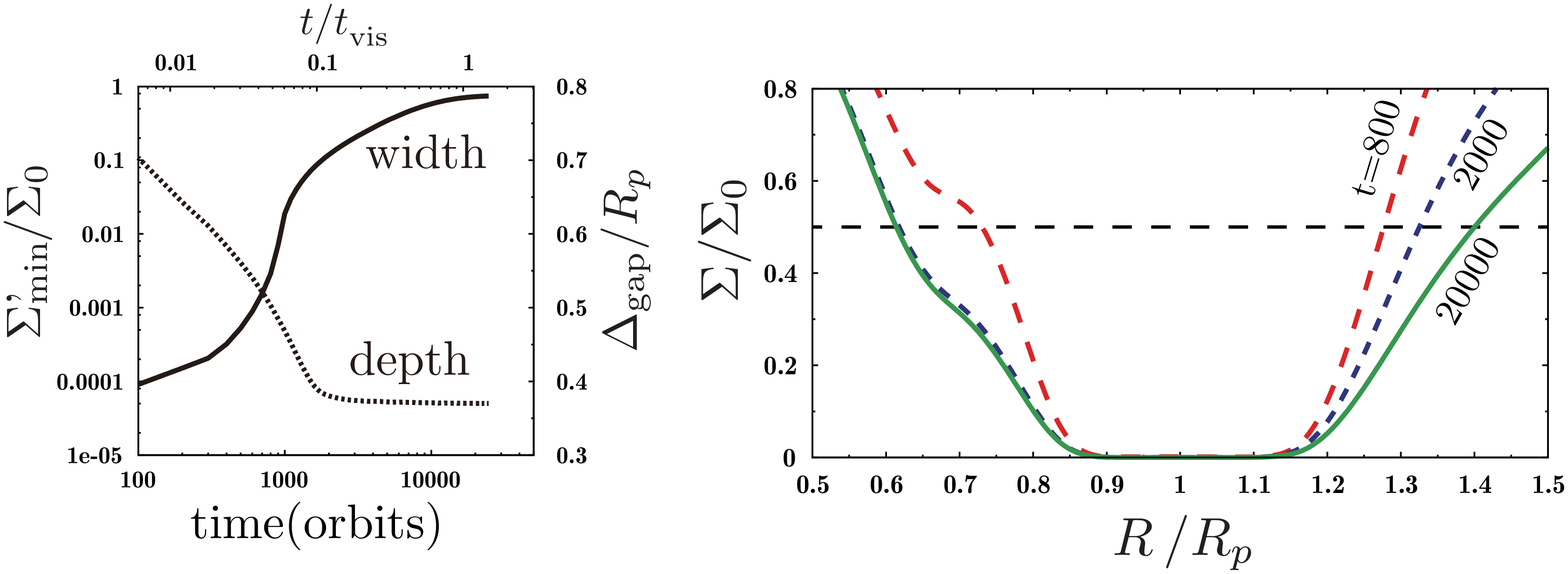}}
		\caption{
		Time variation of the gap structure with $\alpha=10^{-3}$, $\hp/\rp=1/25$, and $M_p/M_{\ast}=10^{-3}$.
		{(\textit{Left})} Time variation of the gap width measured by $\sigmaedge=0.5\Sigma_0$ (solid), and the surface density averaged over the gap bottom (see text;  dotted line).
		({\textit{Right}}) Radial distribution of the azimuthally averaged surface density at $t=800$ (dashed line), $2000$ (dotted line), and $20000$ (solid line) planetary orbits.
		\label{fig:timeevo_a1e-3_h0.04_q1e-3}
		}
	\end{center}
\end{figure*}
Figure~\ref{fig:timeevo_a1e-3_h0.04_q1e-3} shows the time variation of the gap width and depth (left panel) and a snapshot of the azimuthally averaged surface density (right panel), for $\mpl/\mstar=10^{-3}$, $\alpha=10^{-3}$, and $\hp/\rp=1/25$.
We will first consider the time variation of the gap width.
As can be seen in the left panel of Figure~\ref{fig:timeevo_a1e-3_h0.04_q1e-3}, a narrow gap, $\Delta_{\rm gap} = 0.4 R_p$, is formed at $t = 100$ planetary orbits.
The gap width gradually widens with time.
Finally, the gap width reaches $0.79R_p$ at $t\simeq t_{\rm vis} = 1.5 \times 10^{4}$ planetary orbits.
After that, the gap width is almost saturated.
The distribution of the azimuthally averaged surface density becomes wider with time as shown in the right panel.
Note that the evolution of the surface density is slightly asymmetric with respect to the planet; this may be due to the influence of the inner boundary.
However, the gap structure is almost symmetric, as can be seen in the right panel.

\begin{figure*}
	\begin{center}
		\resizebox{0.98\textwidth}{!}{\includegraphics{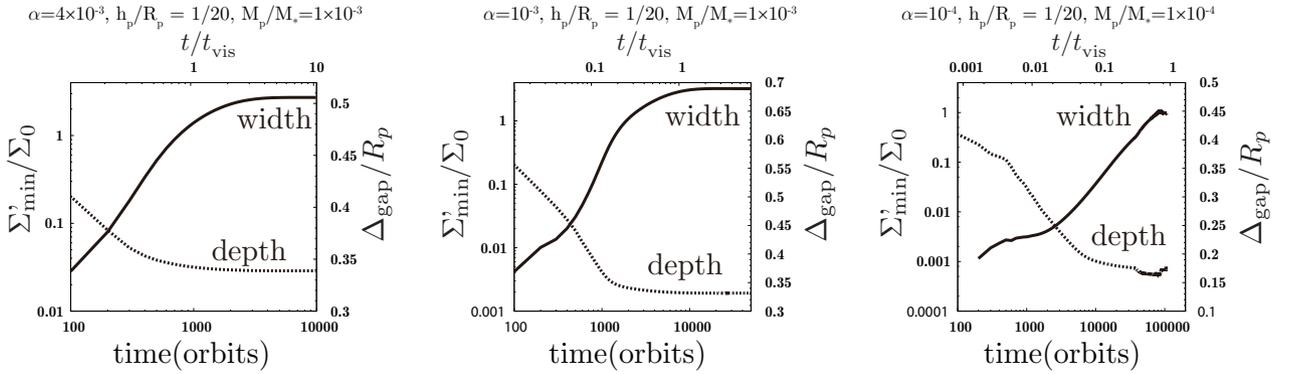}}
		\caption{
 		Time variation of the gap width measured by $\sigmaedge=0.5\Sigma_0$ (solid), and the surface density averaged over the gap bottom.
		In the left panel, $\alpha=4\times 10^{-3}$, $M_{p}/M_{\ast}=10^{-3}$; in the middle panel, $\alpha= 10^{-3}$, $M_{p}/M_{\ast}=10^{-3}$; and in the right panel, $\alpha=10^{-4}$, $M_{p}/M_{\ast}=5 \times 10^{-4}$.
		The disk aspect ratio is set to $1/20$.
		\label{fig:timeevo_various}
		}
	\end{center}
\end{figure*}
In Figure~\ref{fig:timeevo_various}, we show the time variations of the gap width and depth in the cases with different viscosity.
The gap width becomes saturated slower with smaller viscosity, as expected by equation~(\ref{eq:gap_vistime}).
In either cases in the figure, after $t\simeq t_{\rm vis}$, the gap widths are saturated.
The saturated time for the gap depth also depends on the viscosity.
If $\alpha=10^{-4}$ (the right panel of Figure~\ref{fig:timeevo_various}), the gap depth is saturated after $10^4$ orbits, which is about $10$ times larger than that in $\alpha=10^{-3}$ case (the middle panel).
Hence, if the viscosity is very small, a long-term simulation must be required to obtain gaps in steady state.

\section{Comparisons of the gap widths with previous works} \label{sec:width_comp_previous_hydro}
\begin{figure}
	\begin{center}
		\resizebox{0.49\textwidth}{!}{\includegraphics{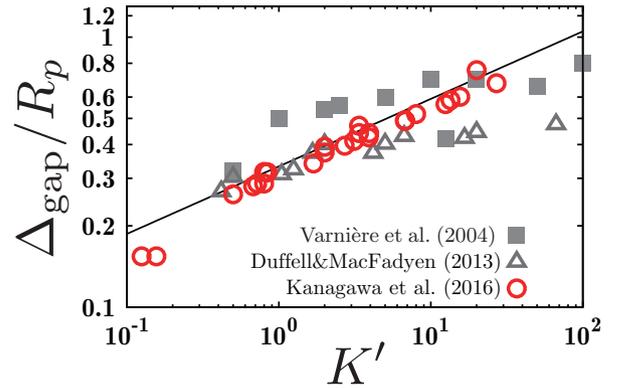}}
		\caption{
		The gap width with $\sigmaedge = (1/3)\Sigma_0$ obtained by simulations done by \protect \cite{Kanagawa2016a} (circles), and the gap widths given by \protect \cite{Duffell_MacFadyen2013} (triangle) and \protect \cite{Varniere_Quillen_Frank2004} (square).
		The thin line indicates the solution to equation~(\ref{eq:gapwidth_varsedge}) with $\sigmaedge=(1/3)\Sigma_0$.
		\label{fig:gapwidth_comp}
		}
	\end{center}
\end{figure}
We compared our result to those of hydrodynamic simulations in previous studies.
Figure~\ref{fig:gapwidth_comp} shows the gap widths given by our simulations and those of \cite{Varniere_Quillen_Frank2004} and \cite{Duffell_MacFadyen2013}.
Because these studies used $\sigmaedge=1/3\Sigma_0$ to define the gap width \footnote{Strictly speaking, \cite{Varniere_Quillen_Frank2004} adopted $\sigmaedge =0.4\Sigma_0$}, we plot the gap widths as measured by $\sigmaedge = 1/3\Sigma_0$.
For $K' \lesssim 10$, the gap widths obtained by previous studies increases as $K'^{1/4}$, which is consistent with the results of equation~(\ref{eq:gapwidth_varsedge}).
As $K'$ continues to increase, the gap widths given by previous studies have a larger scatter respect our results.
In particular, the widths given by \cite{Duffell_MacFadyen2013} are significantly narrower than our results.
They mainly investigated situations with very low viscosity, that is, $\alpha \lesssim 10^{-4}$.
We here remind the reader of the time variations of the gap width described in Appendix~\ref{sec:timeevo_various}.
The time that the gap reaches the saturated width is estimated by the viscous timescale, as in equation~(\ref{eq:vistime_gap}).
When $\alpha=10^{-4}$, the time scale is about $10^{5}$~orbits, although previous studies estimated this to be several thousand planetary orbits.
The variance in the gap width seen in the previous studies can be explained by their short computational times.
Note that when the viscosity is very low, because the time scale of equation~(\ref{eq:vistime_gap}) is very long, the gap cannot reach the saturated width within the lifetime of the disk.
In this case, equation~(\ref{eq:gapwidth_varsedge}) underestimates the planet mass.

\cite{Duffell_Chiang2015} investigated the gap widths in the viscous case ($\alpha \sim 10^{-3}$).
They have measured the gap width using $\sigmaedge=0.1\Sigma_0$ and derived the following empirical formula: $\Delta_{\rm gap}/R_p = 0.5 (\hp/\rp)^{0.22} K'^{0.22}$.
Their empirical formula is quite similar to equation~(\ref{eq:gapwidth_varsedge}) if $\hp/\rp = 0.04$; this is because they have primarily considered cases for which $\hp/\rp \simeq 0.04$.

\section{Model description} \label{sec:model_description}
\subsection{Wave propagation in one-dimensional model}
In order to connect the two-dimensional simulation with the one-dimensional model of K15, we obtain the expression of wave propagation which are given in two-dimensional simulations, for the semianalytical model.
We used the basic equations for the two-dimensional hydrodynamic simulations to derive the basic equation for the one-dimensional model.

Combining the continuity equation and the equation of motion in two-dimensional disk, the equation of angular momentum becomes
\begin{eqnarray}
	&\dpar{R\rhosurf \vphi}{t} + \frac{1}{R}\dpar{\left(R^2 \rhosurf \vphi \vrad \right)}{R}+\dpar{\rhosurf \vphi^2}{\phi}=\nonumber\\
	&\qquad \qquad \qquad \qquad \qquad \qquad -c^2 \dpar{\rhosurf}{\phi}-\rhosurf \dpar{\Psi}{\phi}+R f_{\phi}.
	\label{eq:amf_conv}
\end{eqnarray}
where $f_{\phi}$ is the viscous force per unit area acting in the azimuthal direction, and the gravitational potential is $\Psi$.
We assumed that background structure is axisymmetric and rotates with a constant angular velocity, while the perturbed structure (wave) is nonaxisymmetric and rotates with an angular velocity that depends on $R$ and $\phi$.
Hence, we can decompose the azimuthal velocity into two parts,  the background and the perturbation \citep{Muto_Suzuki_Inutsuka2010}:
\begin{eqnarray}
	\vphi&=\vphiavg(R) + \dvphi(R,\phi).
	\label{eq:decomp_vphi}
\end{eqnarray}
Note that equation~(\ref{eq:decomp_vphi}) is not a linear approximation.
The azimuthally averaged version of equation~(\ref{eq:amf_conv}) can be written as
\begin{eqnarray}
	\dpar{R \overline{\rhosurf \vphi}}{t}+ & \frac{1}{2\pi R}\dpar{\fj}{R}= - \frac{1}{2\pi R} \frac{d\tp}{dR}
	, \label{eq:amfavg_eq}
\end{eqnarray}
where $\fj$ is the azimuthally averaged angular momentum flux,
\begin{eqnarray}
	\fj(R)&=R \vphiavg(R) \fms + \fjwave(R) + \fjvis(R),
	\label{eq:decomp_amf}
\end{eqnarray}
in which where the first term is the angular momentum flux induced by an axisymmetric advection, and the azimuthally averaged mass flux $\fms$ is  $2\pi R \overline{\rhosurf \vrad}$.
The angular momentum fluxes due to nonaxisymmetric advection (waves) $\fjwave$ and viscous diffusion $\fjvis$ are respectively defined by
\begin{eqnarray}
	&\fjwave = 2\pi R^2 \overline{\rhosurf \dvphi \vrad}, \label{eq:amfwave}\\
	&\fjvis  = 2\pi R^3 \overline{\nu \rhosurf \dpar{\Omega}{R}} + 2\pi R \overline{\nu \rhosurf \dpar{\vrad}{\phi}}. \label{eq:amfvis}
\end{eqnarray}
\begin{figure}
	\begin{center}
		\resizebox{0.49\textwidth}{!}{\includegraphics{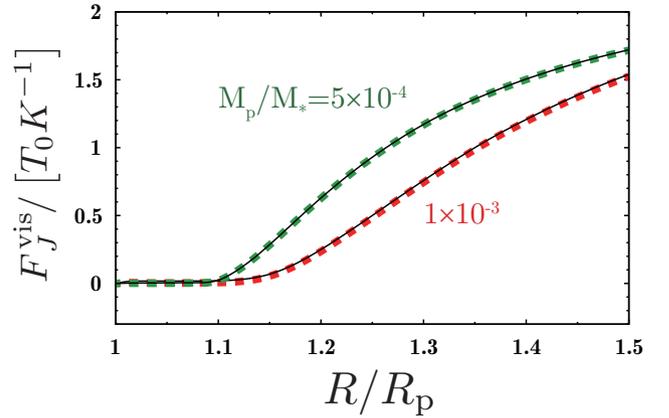}}
	\caption{
	Distribution of the angular momentum flux due to the viscous diffusion for $\mpl/\mstar=5\times 10^{-4}$ and $10^{-3}$ ($\hp/\rp =1/20$ and $\alpha=10^{-3}$).
	The thick dashed lines are the azimuthally averaged flux obtained by equation~(\ref{eq:amfvis}), and the thin solid lines are the azimuthally averaged flux obtained by only the first term of equation~(\ref{eq:amfvis}).
	\label{fig:viscos_torque}
	}
	\end{center}
\end{figure}
In Figure~\ref{fig:viscos_torque}, we show $\fjvis$ and the first term of equation~(\ref{eq:amfvis}) in the cases of $\mpl/\mstar=5 \times 10^{-4}$ and $1\times 10^{-3}$, $\hp/\rp=1/20$ and $\alpha=10^{-3}$.
As shown in the figure, $\fjvis$ is almost equal to the values of the first term of equation~(\ref{eq:amfvis}).
Hence, we can neglect the second term of $\fjvis$ in equation~(\ref{eq:amfvis}).
The right-hand side of equation~(\ref{eq:amfavg_eq}) represents the gravitational torque excited by the planet.
The azimuthally averaged mass flux denoted by $\fms$ is defined to be  $\overline{\rhosurf \vrad}$.
The torque density on the right-hand side of equation~(\ref{eq:amfavg_eq}) is obtained as follows:
\begin{eqnarray}
	\frac{d\tp}{dR}&= 2 \pi R \overline{ \rhosurf \dpar{\Psi}{\phi}},
	\label{eq:tqdens}
\end{eqnarray}
and therefore the torque exerted by the planet is
\begin{eqnarray}
	\tp(R)&= \int^{R}_{\rp} \frac{d\tp}{dR} dR = 2 \pi \int^{R}_{\rp} \overline{ \rhosurf \dpar{\Psi}{\phi}} RdR.
	\label{eq:torque}
\end{eqnarray}
Note that the pressure torque in \cite{Crida_Morbidelli_Masset2006} is given by averaging the first term in RHS of equation~(\ref{eq:amf_conv}) along with stream lines.
Since we azimuhally averaged equation~(\ref{eq:amf_conv}), however, this term is vanished.
Instead of the pressure torque, we consider the angular momentum flux due to the waves described by equation~(\ref{eq:amfwave})
.
To clarify the effect of the wave propagation, equation~(\ref{eq:amfavg_eq}) can be rewritten as
\begin{eqnarray}
	&\dpar{R \overline{\rhosurf \vphi}}{t}+ \frac{1}{2\pi R}\dpar{\left(R\vphiavg\fms\right)}{R} +  \frac{1}{2\pi R}\dpar{\fjvis}{R} = \nonumber \\
	 &\qquad \qquad \qquad \qquad - \frac{1}{2\pi R} \left( \frac{d\tp}{dR}-\frac{d\fjwave}{dR} \right).
	\label{eq:amfavg_eq2}
\end{eqnarray}
First, the planet excites the waves and thus transfers angular momentum to them.
Next, this angular momentum is deposited onto the disk as the waves are damped.
Because of this wave propagation, the deposition rate of angular momentum to the disk gas is different from the torque density excited by the planet.
The deposition rate from the waves is given by
\begin{eqnarray}
	\lambdad(R) &= \frac{d\tp}{dR} - \frac{d\fjwave}{dR}.
	\label{eq:deposition_rate_def}
\end{eqnarray}
The cumulative deposited torque is given by
\begin{eqnarray}
	\tdeposit(R) &= \int^{R}_{\rp} \Lambda_d (R')dR' = \tp(R)-\fjwave(R).
	\label{eq:def_deposited_amf}
\end{eqnarray}

Integrating equation~(\ref{eq:amfavg_eq}), we obtain an equation for the angular momentum flux, as follows:
\begin{eqnarray}
	&R \vphiavg \fms  +  \fjvis  =\qquad \qquad \qquad \qquad \nonumber\\
	&\qquad \quad  \fj(\rp) + \tp - \fjwave - \int^{R}_{\rp} \dpar{R^2\overline{\rhosurf \vphi}}{t} R dR.
	\label{eq:flux}
\end{eqnarray}
In steady state, the last term on the right-hand side of equation~(\ref{eq:flux}) should be zero.
Hence, we finally obtain
\begin{eqnarray}
	R \vphiavg \fms  + \fjvis (R)= \fj(\rp) + \tdeposit(R),
	\label{eq:flux2a}
\end{eqnarray}
which is the same as the basic equation (equation~9) of K15.

\subsection{Model of wave excitation} \label{subsec:model_waveexcitation}
In the semianalytical model, we adopt the WKB formula for the planetary torque density \citepeg{Ward1986} and use a simple cutoff, as in K15:
\begin{eqnarray}
	\frac{d\tp}{dR}&= \left\{
	\begin{array}{l}
		\pm 0.40 f_{\rm NL} {\displaystyle \frac{\rhosurf(R)}{\rhosurf_0}  \left( \frac{\hp}{R-\rp} \right)^{4} \left( \frac{d\tp}{dR} \right)_0}\\
		\qquad \qquad \qquad \qquad  \mbox{for } |R-\rp| > \Delta,\\
		0  \qquad \qquad \qquad \quad \mbox{for } |R-\rp| \leq \Delta,
	\end{array}\right.
	\label{eq:wkbtorque}
\end{eqnarray}
where
\begin{eqnarray}
	\left( \frac{d\tp}{dR} \right)_0 &= 2 \pi \rp^3 \Omega^{2}_{p} \Sigma_{0} \left( \frac{\mpl}{\mstar} \right)^{2} \left( \frac{\hp}{\rp} \right)^{-4}.
	\label{eq:coef_tqdens}
\end{eqnarray}
For the $\pm$ sign in equation~(\ref{eq:wkbtorque}), $+$ is taken for $R>\rp$ and $-$ is taken for $R<\rp$.
The cutoff length $\Delta$ should set to $1.3\hp$ to match the one-side torque obtained from  linear analyses of the disk--planet interaction \citep{Takeuchi_Miyama1998,Tanaka_Takeuchi_Ward2002,Muto_Inutsuka2009}.
The factor $f_{\rm NL}$ represents the reduction rate of the excitation torque due to the nonlinear effect.

\RED{
The value of $f_{\rm NL}$ should be unity when the excitation of waves is in linear regime as $\mpl/\hp^3 < 1$.
When $\mpl/\hp^3 > 1$, $f_{\rm NL}$ should take a value smaller than unity, since the excitation of waves is in non-linear regime.
Using $l=\mpl/\hp^3$, the parameter $K$ is expressed as $l^2\hp/\alpha$.
When $l > 1$, $K > \hp/\alpha$.
In the parameter space we searched ($1/30 < \hp/\rp < 1/15$, $10^{-4} < \alpha < 10^{-2}$), the excitation of waves is nonlinear when $K\sim 10^2$.
Although the transition from the linear regime to the non-linear regime should be smooth, a step function as $f_{\rm NL}=1$ when $\mpl/\hp^3 < 1$, and $f_{\rm NL}=0.4$ when $\mpl/\hp^3>1$ would be acceptable because the depth around $K\simeq 10^{2}$ does not sensitively depend on $f_{\rm NL}$ as seen from the left panel of Figure~\ref{fig:gapdepth_comp}.
}

The one-side torque exerted by the planet is given by
\begin{eqnarray}
	\tponeside&=\tp(\infty)=2\pi \int^{R\rightarrow\infty}_{\rp} \overline{\rhosurf \dpar{\Psi}{\phi}} R'dR'.
	\label{eq:one-sided_torque}
\end{eqnarray}
 Using equation~(\ref{eq:wkbtorque}), we obtain $\tponeside$ as
\begin{eqnarray}
\tponeside &= 0.4 f_{\rm NL} \left(\frac{d\tp}{dR}\right)_0 \int^{\infty}_{\rp+\Delta} \frac{\Sigma(R)}{\Sigma_0} \left(\frac{\hp}{R-\rp} \right)^{4} dR.
\label{eq:one-sided_torque2}
\end{eqnarray}
The one-side torque $\tponeside$ depends on the surface density, as seen in equation~(\ref{eq:one-sided_torque2}).
Hence, when we solve equation~(\ref{eq:flux2a}),  iteration is necessary in order to consistently obtain $\rhosurf$ and $\Lambda_d$ (or $\tdeposit$).
For details, see Section~2.4 in K15.

\RED{
\section{Condition of the Rossby wave instability in the low viscous case} \label{sec:RWI_inviscid}
\begin{figure}
	\begin{center}
		\resizebox{0.48\textwidth}{!}{\includegraphics{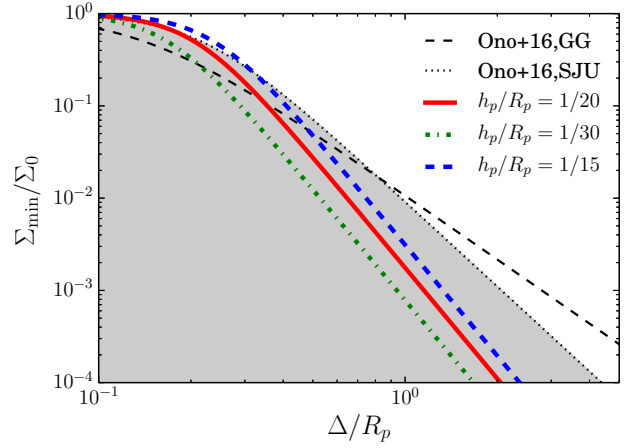}}
	\caption{
	\label{fig:RWI_condition}
	The gap depth-width relationship given by equation~(\ref{eq:rel_depth_width}) with $\sigmaedge=0.5\Sigma_0$ for $\hp/\rp=1/20$ (solid), $1/30$ (dot-dash), and $1/15$ (dashed), respectively.
	The vertical and horizontal axes indicate the minimum surface density of the gap and the full width of the gap, respectively.
	In the gray-filled region, the RWI is unstable in the case of SJU structure.
	}
	\end{center}
\end{figure}
\cite{Ono2016} have obtained the necessary and sufficient condition for the onset of the RWI, for various structures in the inviscid case.
They have derived the threshold minimum surface density when the RWI is marginal stable, as a function of the width of the structures.
To discuss the RWI in the gap, we compare the relations of the threshold minimum density and the width for a Gaussian gap (GG) and a step jump-up (SJU) structures (which approximately models the gap structure outer than the planetary orbit) to the depth-width relation of the gap given by equation~(\ref{eq:rel_depth_width}).
According to \cite{Ono2016}, the threshold minimum density of the RWI is obtained by $\sigmamin/\Sigma_0|_{\rm crit} = [1+a(\Delta/\rp)^{b}]^{-1}$, where $\Delta$ is the full width of the gap, and the parameter $a$ and $b$ are given by $93.1$ and $2.31$ for the Gaussian gap, and $108.6$ and $3.06$ for the step jump-up structure\footnote{\cite{Ono2016} adopted the parameter A, instead of $\sigmamin/\Sigma_0$. Using the parameter A, $\sigmamin/\Sigma_0=1/(1+A)$. In the case of the step jump-up structure, we adjust the value of the parameter $a$ from Table~2 of \cite{Ono2016} since we treat the threshold minimum density as a function of the full width of the gap (\cite{Ono2016} treated the threshold density as a function of the half width of the gap ($=\Delta/2$)).}.
If the gap depth is smaller than the threshold minimum surface density of the RWI, the vortex would be formed by the RWI.
Note that \cite{Ono2016} investigated the RWI in relatively narrow gaps ($\Delta/\rp < 0.2$), since the RWI is stabilized in the wide gap as $\Delta \gg \hp$.
Although we apply their formula to a wide gap, the RWI is possible to be stable even if the gap minimum density is smaller than the threshold density given by Ono's criterion when $\Delta/\rp$ is large.
In Figure~\ref{fig:RWI_condition}, we show the depth-width relation of the gap which is adopted by $\sigmaedge=0.5\Sigma_0$, and the threshold minimum surface density of the RWI for the Gaussian gap and the step jump-up function.
As shown in the figure, the RWI can occur in the planet-induced gap, if the viscosity is very low, which is consistent with the previous studies \citepeg{Yu2010,Fu_Li_Lubow_Li2014,Zhu_Stone_Rafikov_Bai2014}.
If the RWI occurs, the gap structure in steady state may be determined by the marginal condition of the RWI.
In this case, the depth-width relation of the gap may be determined by Ono's criterion, instead of our relation of equation~(\ref{eq:rel_depth_width}).
}

\end{document}